\documentclass[aps,preprintnumbers,twocolumn,amsmath,amssymb,floatfix,superscriptaddress,pra]{revtex4-1}
\pdfoutput=1
\usepackage[utf8]{inputenc}
\usepackage{amssymb}
\usepackage{amsmath,bm,bbm}
\usepackage{amsfonts}
\usepackage{braket}
\usepackage{tensor}
\usepackage{graphicx}
\usepackage[colorlinks=true,linkcolor=blue, citecolor=blue, urlcolor=blue, bookmarks]{hyperref}
\usepackage[normalem]{ulem}

\def\be{\begin{equation}}
\def\ee{\end{equation}}
\def\bea{\begin{eqnarray}}
\def\eea{\end{eqnarray}}
\def\bes{\begin{subequations}}
\def\esu{\end{subequations}}

\def\XXint#1#2#3{{\setbox0=\hbox{$#1{#2#3}{\int}$}
         \vcenter{\hbox{$#2#3$}}\kern-.5\wd0}}

\newcommand\la		{\lambda}
\renewcommand\th          {\theta}
\newcommand{\ud}          {\mathrm d}

\newcommand \rhoh       {\rho^{\text{h}}}
\newcommand \rhot       {\rho^{\text{t}}}
\newcommand\w           {\omega}
\newcommand\limsc     {{\textstyle \lim_{\rm NR}}}

\newcommand{\1}{\mathbbm{1}}

\begin{document}

\title{Transport in the sine--Gordon field theory:\\
from generalized hydrodynamics to semiclassics}

\author{Bruno Bertini}
\affiliation{Department of Physics, Faculty of Mathematics and Physics, University of Ljubljana, Jadranska 19, SI-1000 Ljubljana, Slovenia}
\author{Lorenzo Piroli}
\affiliation{Max-Planck-Institut f\"ur Quantenoptik, Hans-Kopfermann-Str. 1, 85748 Garching, Germany}
\author{M\'arton Kormos}
\affiliation{BME ``Momentum" Statistical Field Theory Research Group
	1111 Budapest, Budafoki \'ut 8, Hungary}
\affiliation{Department of Theoretical Physics,
	Budapest University of Technology and Economics
	1111 Budapest, Budafoki \'ut 8, Hungary}

\begin{abstract}
The semiclassical approach introduced by Sachdev and collaborators proved to be extremely successful in the study of quantum quenches in massive field theories, both in homogeneous and inhomogeneous settings. While conceptually very simple, this method allows one to obtain analytic predictions for several observables when the density of excitations produced by the quench is small. At the same time, a novel generalized hydrodynamic (GHD) approach, which captures exactly many asymptotic features of the integrable dynamics, has recently been introduced. Interestingly, also this theory has a natural interpretation in terms of semiclassical particles and it is then natural to compare the two approaches. This is the objective of this work: we carry out a systematic comparison between the two methods in the prototypical example of the sine--Gordon field theory. In particular, we study the ``bipartitioning protocol'' where the two halves of a system initially prepared at different temperatures are joined together and then left to evolve unitarily with the same Hamiltonian. We identify two different limits in which the semiclassical predictions are analytically recovered from GHD: a particular non-relativistic limit and the low temperature regime. Interestingly, the transport of topological charge becomes sub-ballistic in these cases. Away from these limits we find that the semiclassical predictions are only approximate and, in contrast to the latter, the transport is always ballistic. This statement seems to hold true even for the so-called ``hybrid'' semiclassical approach, where finite time DMRG simulations are used to describe the evolution in the internal space.
\end{abstract}

\maketitle

\section{Introduction}
\label{sec:intro}

Semiclassical approaches have accompanied the study of quantum mechanics ever since its first formulations \cite{Dirac_book}, and nowadays represent a standard toolbox for theoretical physicists. These approaches allow one to simplify the description of quantum systems by treating classically some aspects of their dynamics, and are of interest even when more sophisticated tools are available. Consider, for instance, the case of one-dimensional integrable models with internal degrees of freedom~\cite{Sutherland_book}. While these systems are amenable to exact analyses, such as the nested Bethe ansatz \cite{Korepin_book}, some of their properties remain generically hard to determine, most notably correlation functions. In this case, Sachdev and collaborators~\cite{Sachdev_book, SaYo97, DaSa05,DaSa98,SaDa97} showed that the low temperature correlators can be determined by an intuitive semiclassical picture formulated in terms of quasiparticles propagating through the system [see also~\cite{RZ:semiclassics, RaZa06, RaZa09}]. 

Semiclassical approaches are even more relevant in the study of non-equilibrium physics \cite{CaEM16}. 
Indeed, the full description of the out-of-equilibrium dynamics is notoriously hard, even in simple protocols such as quantum quenches \cite{cc-06}, and semiclassical approaches often represent the only option to obtain analytical insight~\cite{RiIg11, RSMS09, Evan13, KoZa16, IgRi11,DiIR11,RiIg11,BlRI12}. It is then crucial to understand their precise range of applicability, testing them against exact results coming from integrability in instances where the latter are available. In this respect, ${1+1}$ dimensional integrable quantum field theories are of particular interest. Indeed, they offer powerful tools coming from the combination of integrability, relativistic invariance, and non-trivial analytical structures in momentum space~\cite{mussardo_book}, for instance form-factor expansions \cite{smirnov_book, Delf04}. 

Specifically, an ideal testing ground where both semiclassical and integrability based methods can be applied is given by the sine--Gordon field theory. Apart from being the prototypical example of an integrable quantum field theory with an internal $O(2)$ symmetry, this theory also attracts a large amount of attention from the condensed matter community, mainly because it provides a description of the low-energy physics of several spin chains~\cite{EsKo05}, and of interacting one dimensional bosons~\cite{Giamarchibook, GPD:coupledcondensates}.

The quench dynamics of the sine--Gordon field theory was first investigated in \cite{BeSE14} by combining form factor expansions with the Quench Action method \cite{CaEs13, Caux16,DWBC14}, an exact approach deeply rooted in the theory of integrability. There, the authors considered homogeneous settings and determined the full dynamics of a particular vertex operator in the limit of small energy densities. Remarkably, the results of \cite{BeSE14} were later exactly recovered in \cite{KoZa16} by a generalisation of Sachdev's semiclassical approach, revealing all its appeal also out of equilibrium. Furthermore, it was shown that this semiclassical approach can be developed further into a hybrid method where, at first order, quantum effects could be taken into account either via Monte Carlo simulations or via a complete quantum mechanical treatment of internal degrees of freedom \cite{MoKZ17,WMLKZ19}.

More recently, increasing interest has been devoted to the study of inhomogeneous situations. A particularly simple, yet non-trivial, setting which has been widely investigated is the so-called bipartitioning protocol, where two semi-infinite systems, initially held at different temperatures, are suddenly joined together and left to evolve unitarily under a homogeneous Hamiltonian. While this protocol is of obvious interest for the study of transport, until recently its analytic understanding was limited to the cases of free \cite{ARRS99,AsPi03,AsBa06,PlKa07,LaMi10,EiRz13,DVBD13,Bert17,ADSV16, CoKa14,EiZi14,CoMa14,DeMV15,DLSB15,VSDH16,KoZi16,Korm17,PeGa17} or conformally invariant models \cite{SoCa08,CaHD08,Mint11,MiSo13,DoHB14,BeDo12,BeDo15,BeDo16,BeDo16Review,LLMM17,DuSC17, SDCV17}. A major breakthrough came in Refs.~\cite{CaDY16,BCDF16}, with the introduction of the theory of generalized hydrodynamics (GHD) which allows one to treat transport even in \emph{interacting} integrable models. Within this approach, at large times the system is described by a family of space-dependent quasi-stationary states, determined by a set of continuity equations for the densities of conserved charges. 

Bipartitioning protocols have been investigated in detail using GHD in different models, ranging from spin chains \cite{PDCB17,BVKM17,Alba18,Alba19,CoDV18,BePi18,BePC18,BFPC18,DeBD18,MVCR18,DMKI19,GoVa18,BVKM18} and electronic systems \cite{IlDe17,IlDe17_II}, to non-relativistic quantum gases \cite{DDKY17,DoSp17,CDDK17,MBPC19} and classical field theories \cite{BDWY18}, unveiling a rich phenomenology in the transport of one-dimensional integrable theories. Remarkably, it was found in \cite{DoYC18} that the GHD equations are in a sense classical: for a given quantum system one can find quite generally a classical counterpart which is described by the same set of hydrodynamic equations in the limit of large space and time scales. In light of this, it is natural to wonder what the relation between GHD and Sachdev's semiclassical approach is. Investigating such relation is the purpose of this paper. Our strategy is to compare their predictions  for the transport generated by bipartitioning protocols in the sine--Gordon model.

A semiclassical analysis of bipartitioning protocols has already been performed in \cite{KoMZ18} and a number of interesting qualitative features have been revealed. Most prominently, Ref.~\cite{KoMZ18} pointed out that the transport of the charge associated with the internal degree of freedom is always diffusive. In essence, this is due to the fact that each semiclassical particle carries a discrete unit of the internal charge and two semiclassical particles originated close to each other can not get too far due to the interactions with the other particles; this bounds the spreading of the charge. Note that GHD can also predict similar sub-ballistic transport, e.g., in the case of gapped XXZ spin-1/2 chains~\cite{PDCB17}. This happens when the GHD particles responsible for the transport of a certain charge become non-dispersive, i.e., they all move with a single velocity.

Our results can be summarised as follows. First, we show that the semiclassical predictions for the profiles of local observables can be analytically recovered from GHD in two cases: (i) low temperatures and (ii) a particular non-relativistic limit. The first case is very close to the setting where Sachdev's semiclassical approach was first introduced, and corresponds to a small density of excitations over the ground state. In the second case the limit is taken in such a way that the assumptions of the semiclassical theory are exactly fulfilled by the quantum dynamics, so that the semiclassical predictions become exact even at finite densities. Away from these regimes, however, we find that the semiclassical predictions are only approximate. In particular, we find that, contrary to the semiclassical prediction, the charge associated with the internal symmetry [topological charge] always spreads ballistically. This seems not to be captured even by more refined versions of the semiclassical approach, such as the ``hybrid'' version of Ref.~\cite{MoKZ17}: as we discuss later on, we ascribe this to the fact that the semiclassical picture is formulated in terms of the ``wrong'' quasiparticles.

The rest of the paper is organised as follows. In Sec.~\ref{sec:model} we briefly introduce the sine--Gordon field theory and its Thermodynamic Bethe Ansatz (TBA) description. In Sec.~\ref{sec:bipartitioning} we summarise semiclassical and GHD descriptions of the transport generated by the bipartitioning protocol. In Sec.~\ref{sec:comparisonlim} we show how the semiclassical predictions are analytically recovered from GHD in the aforementioned limiting cases. In Sec.~\ref{sec:comparisongen} we compare the two approaches for generic temperatures and couplings. Finally, Sec.~\ref{sec:conclusions} contains our conclusions. A number of appendices complements the main text with more technical aspects.

\section{The sine--Gordon field theory}
\label{sec:model}

\subsection{Hamiltonian and $S$-matrix}
\label{sec:hamandS}

The sine--Gordon field theory is a (1+1)-dimensional bosonic quantum field theory described by the following Hamiltonian
\be
\!\!H =\!\!\! \int \!\!{\rm d}x \!\left[\frac{c^2}{2} :\!\Pi^2\!: +\frac{1}{2} :\!(\partial_x\Phi)^2\!:\! -\frac{c^2 \alpha^2}{\beta_{\rm sG}^2} :\!\cos(\beta_{\rm sG}\Phi)\!\!:\right]\!,
\label{eq:ham}
\ee
where the real bosonic field $\Phi$ and its conjugate momentum $\Pi$ satisfy the canonical commutation relations 
\be
[\Phi(x),\Pi(y)]=i\delta(x-y)\,.
\ee 
In \eqref{eq:ham} $:\cdots:$ denotes normal ordering with respect to the physical ground state, the parameter $\alpha$ gives a mass scale, the parameter $\beta_{\rm sG}$ is the coupling, and $c$ is the speed of light [and we set $\hbar=1$]. Note that, without loss of generality, we can choose ${\beta_{\rm sG} \geq 0}$. Moreover, the Hamiltonian \eqref{eq:ham} is bounded from below, and thus physically meaningful, only for ${\beta_{\rm sG}\leq \sqrt{8\pi/c}}$~\cite{Cole75}.  

The field theory defined by \eqref{eq:ham} is integrable, i.e., it features an infinite number of local conservation laws (or charges) $ \{I_n\}_{n=1,2,...}$ which constrain the dynamics. This set includes the momentum $I_1=P$, the Hamiltonian itself $I_2=H$, together with the ``higher'' conservation laws $I_{n>2}$. The constraints imposed by the latter are most easily described by viewing \eqref{eq:ham} as a scattering theory of relativistic particles corresponding to excitations above a vacuum [this is always possible for field theories in Minkowski space]. The presence of the set $ \{I_n\}_{n=1,2,...}$ forces the scattering of such particles to be purely elastic and completely factorisable in terms of two-body processes~\cite{mussardo_book}. An immediate consequence of this is that the number $N$ of such particles is conserved; it is, however, not an independent charge as its conservation is ``encoded" in the set $\{I_n\}_{n=1,2,...}$.

Together with the charges $ \{I_n\}_{n=1,2,...}$ the theory also features two important internal symmetries. First of all, it is immediate to see that the Hamiltonian \eqref{eq:ham} is left invariant by the following $\mathbb Z_2$ symmetry transformation
\be
\Phi(x)\mapsto R\Phi(x)=-\Phi(x)\,.
\ee
Moreover, even if it is not apparent from \eqref{eq:ham}, the theory also displays an additional $O(2)$ symmetry, which becomes manifest in the fermionic formulation of the model~\cite{Cole75}, where it corresponds to the conservation of electric charge. The conserved quantity associated with this continuous symmetry is called ``topological charge'' and is defined as 
\be
Q=\frac{\beta_{\rm sG}}{2\pi}\int{\rm d}x\,\,\partial_x\Phi\,.
\label{eq:tcharge}
\ee
This $O(2)$ symmetry is also explicitly displayed by the eigenstates of \eqref{eq:ham} as we now briefly review.   

The spectrum of the model for ${\beta_{\rm sG}\in[0,\sqrt{8\pi/c}]}$ is conveniently parametrised in terms of the ``renormalised'' coupling~\cite{rajaraman_book} 
\be
p=\frac{c\beta_{\rm sG}^2}{8\pi-c\beta_{\rm sG}^2}\,.
\ee
For $p\geq1$ the system is in the ``repulsive regime'' and the eigenstates of the Hamiltonian can be viewed as scattering states of two elementary relativistic particles of equal mass $M$ and opposite charge $\pm1,$ called ``soliton'' and ``antisoliton''. The expression of the mass $M$ in terms of the Hamiltonian's parameters has been determined in Ref.~\cite{Zamolodchikov:massscale}; in our conventions it reads
\be
M=\sqrt\frac{4 \cot(\frac{\pi p}{2})}{(1+p)c}\frac{\alpha}{\beta_{\rm sG}}\,.
\label{eq:solitonmass}
\ee
For $p\in[0,1[$, instead, the system is in the so-called ``attractive regime'', where solitons and antisolitons can form bound states called ``breathers''. Specifically,  for a given $p\in[0,1[$ the spectrum features $\lfloor1/p\rfloor$ different breathers with masses given in terms of $M$ as follows  
\be
M_k=2M \sin\left(\frac{\pi k p}{2}\right),\qquad k=1,\ldots,\lfloor1/p\rfloor\,.
\ee
The point $p=1$ is special, as for this value of $p$ the theory is equivalent to a free massive Dirac fermion. 

Note that the appearance in the spectrum of solitons and antisolitons is a non-perturbative effect: these particles cannot be observed in perturbation theory in $\beta_{\rm sG}$. For instance, this can be seen by expanding \eqref{eq:solitonmass} for small $\beta_{\rm sG}$ and verifying that the expansion is not written in terms of a power series. Indeed, the explicit calculation yields 
\be
M=\frac{8\alpha}{c \beta_{\rm sG}^2}+O(\beta_{\rm sG}^0)\,.
\ee
The ``perturbative'' particle associated to the field $\Phi$ for small $\beta_{\rm sG}$ is the first breather. Consistently, its mass has the following expansion for small $\beta_{\rm sG}$
\be
M_1=\alpha+O(\beta_{\rm sG})\,.
\ee 

In this work we always consider the non-perturbative regime $p\geq1$ where the field $\Phi$ does not create any single-particle excitation. In this case, a basis of eigenstates can be conveniently constructed by means of the following Faddeev--Zamolodchikov creation and annihilation operators~\cite{Faddeev, ZZ:factorisedSmatrices} 
\be
Z^\dagger_a(\theta),\,\, Z_a(\theta)\,,\qquad a=\pm,\quad\theta\in \mathbb R\,.
\label{eq:Zop}
\ee
These operators are interpreted as creation and annihilation operators for solitons [$a=+$] and antisolitons [$a=-$] with rapidity $\theta$, and obey the following algebra 
\bes
\begin{align}
Z_{a_1}(\theta_1)Z_{a_2}(\theta_2)&=
S_{a_1a_2}^{b_1b_2}(\theta_1-\theta_2)Z_{b_2}(\theta_2)Z_{b_1}(\theta_1)\,,\\
Z^\dagger_{a_1}(\theta_1)Z^\dagger_{a_2}(\theta_2)&=
S_{a_1a_2}^{b_1b_2}(\theta_1-\theta_2)Z^\dagger_{b_2}(\theta_2)Z^\dagger_{b_1}(\theta_1)\,,\\
Z_{a_1}(\theta_1)Z^\dagger_{a_2}(\theta_2)&=2\pi\delta(\theta_1-\theta_2)
\delta_{a_1,a_2}\notag\\
&\,\,\,\,\,\,+S_{a_2b_1}^{b_2a_1}(\theta_2-\theta_1)Z^\dagger_{b_2}(\theta_2)Z_{b_1}(\theta_1)\,.
\end{align}
\esu
Here the sum over repeated indices is implicit and we denoted by $S_{ab}^{cd}(\theta)$ the two-particle scattering matrix of the sine--Gordon model in the repulsive regime. The non-zero elements of $S_{ab}^{cd}(\theta)$ read as~\cite{ZZ:factorisedSmatrices, KorepinSG} 
\bes
\label{eq:Smatrix}
\begin{align}
S^{++}_{++}(\theta)&=S^{--}_{--}(\theta)\equiv S_0(\theta),\\
\frac{S^{+-}_{+-}(\theta)}{S_0(\theta)}&=\frac{S^{-+}_{-+}(\theta)}{S_0(\theta)} = \frac{\sinh\big(\frac{\theta}{p}\big)}
{\sinh\big(\frac{ i \pi-\theta}{p}\big)}  \equiv S_{T}(\theta),\label{eq:ST}\\
\frac{S^{+-}_{-+}(\theta)}{S_0(\theta)}&=\frac{S^{-+}_{+-}(\theta)}{S_0(\theta)}= \frac{i \sin\big(\frac{\pi}{p}\big)}
{\sinh\big(\frac{i\pi-\theta}{p}\big)}\equiv S_{R}(\theta)\label{eq:SR},
\end{align}
\esu
where we introduced
\be
\!\!\!S_0(\theta)=
\!-\exp\!\left[i\!\! \int_0^\infty\!\!\frac{dt}{t}
\sin\!\!\left(\frac{t\theta}{\pi p}\right)\frac{\sinh\big(\frac{p-1}{2p}t\big)}
{\sinh\big(\frac{t}{2}\big)\cosh\big(\frac{t}{2p}\big)}\!\right]\!\!.
\label{eq:S0}
\ee
The S-matrix fulfils the Yang--Baxter equation together with crossing and unitarity relations
\bes
\begin{align}
&S^{b_2b_3}_{a_2a_3}(\theta_2-\theta_3)S_{a_1b_3}^{b_1c_3}(\theta_1-\theta_3)S_{b_1b_2}^{c_1c_2}(\theta_1-\theta_2)\notag\\
&=S_{a_1a_2}^{b_1b_2}(\theta_1-\theta_2)S_{b_1a_3}^{c_1b_3}(\theta_1-\theta_3)S_{b_2b_3}^{c_2c_3}(\theta_2-\theta_3), 
\label{eq:SYB}\\
&S_{a_1a_2}^{c_1c_2}(\theta)S_{c_1c_2}^{b_1b_2}(-\theta)=
\delta_{a_1}^{b_1}\delta_{a_2}^{b_2},
\label{eq:Sunitarity}\\
&S_{ab}^{cd}(i\pi-\theta)=S_{\bar{c}b}^{\bar{a}d}(\theta)
=S_{a\bar{d}}^{c\bar{b}}(\theta),\quad \bar{a}=-a\,.
\label{eq:Scrossing}
\end{align}
\esu
Moreover, we have 
\be
S_{a_1a_2}^{b_1b_2}(0)=-\delta_{a_1}^{b_2}\delta_{a_2}^{b_1}\,.
\label{eq:zeromomentum}
\ee
In this framework, the ground state $|0\rangle$ is defined as the state annihilated by all the Faddeev--Zamolodchikov annihilation operators, 
\be
Z_a(\theta)|0\rangle=0\,. 
\ee
With these ingredients, a basis of eigenstates of $H$ is obtained as follows 
\be
\!\!\mathcal B=\{\ket{\boldsymbol \theta}_{\boldsymbol a}:\, \theta_j\!\in \!\mathbb R,\, a_j\!\in\!\!\{\pm\},\,\theta_j<\theta_{j+1}\},
\label{eq:scatteringstates}
\ee
where ${\boldsymbol \theta=(\theta_1,\ldots,\theta_N)}$ and ${\boldsymbol a=(a_1,\ldots,a_N)}$ are $N$-component vectors, while we introduced 
\be
\ket{\boldsymbol \theta}_{\boldsymbol a}\equiv Z^\dagger_{a_1}\!(\theta_1)\ldots
Z^\dagger_{a_N}\!(\theta_N)|0\rangle\,.
\label{eq:scatteringstate}
\ee
The states \eqref{eq:scatteringstates} simultaneously diagonalise the infinite set of commuting conserved charges in involution related to the integrability of the model. Moreover, they also diagonalise the topological charge \eqref{eq:tcharge}. In particular their eigenvalues of energy, momentum, and topological charge read as
\bes
\begin{align}
&\tensor[_{\boldsymbol a}]{\braket{\boldsymbol \theta| H |\boldsymbol \theta}}{_{\boldsymbol a}}= \sum_{j=1}^N M c^2 \cosh\theta_j,\label{eq:energyscatteringstates}\\
&\tensor[_{\boldsymbol a}]{\braket{\boldsymbol \theta| P |\boldsymbol \theta}}{_{\boldsymbol a}}=\sum_{j=1}^N M c \sinh\theta_j,\label{eq:momentumscatteringstates}\\
&\tensor[_{\boldsymbol a}]{\braket{\boldsymbol \theta| Q |\boldsymbol \theta}}{_{\boldsymbol a}}=\sum_{j=1}^{N}a_j.\label{tcscatteringstates}
\end{align}
\esu

\subsection{Eigenstates in finite volume}

Up to now we described the properties of the sine--Gordon field theory in infinite volume. To treat finite densities of particles, however, we need to consider the theory confined in a finite volume $L$ and take the thermodynamic limit, i.e., $L,N\rightarrow\infty$ with a finite ratio $D=N/L$. This situation is conveniently analysed within the framework of the Thermodynamic Bethe Ansatz (TBA) for relativistic integrable quantum field theories~\cite{ZamoTBA}, which we briefly review in the following. 

The leading effect of confining the theory in a finite volume is that the rapidities characterising a given eigenstate of the Hamiltonian become quantised. In other words, a given eigenstate of $H$ is allowed in finite volume only if its rapidities fulfil some conditions depending on $L$. In general, since the theory is nontrivially interacting, the quantisation conditions couple together all the rapidities of the state. To be more specific, let us consider an eigenstate of the Hamiltonian specified by the rapidities $\boldsymbol \theta=\{\theta_k\}_{k=1}^N$ and impose periodic boundary conditions. In this case the allowed eigenstates are those in the kernel of the operators
\be
{\cal O}_j(\boldsymbol \theta)= {\cal T}(\theta_j|\boldsymbol \theta)-e^{-i L M c \sinh \theta_j} \1 ,\quad j=1,\ldots,N,
\label{eq:boundarycondop}
\ee
where $\1$ is the $2^N \times 2^N$ identity and $ {\cal T}(\lambda|\boldsymbol \theta)$ is the $N$-particle transfer matrix, defined through its matrix elements in the basis \eqref{eq:scatteringstates} as  
\be
{\cal T}(\lambda|\boldsymbol \theta)^{\boldsymbol b}_{\boldsymbol a}=S_{c_N a_1}^{c_1b_1}(\lambda-\theta_1)\cdots S_{c_{N-1} a_N}^{c_N b_N}(\lambda-\theta_N)\,.
\ee
The condition of being in the kernel of the operators~\eqref{eq:boundarycondop} can be interpreted as the requirement that subsequently swapping any particle in the state with all the others is the same as translating it by $L$. As the transfer matrix is nondiagonal in the basis \eqref{eq:scatteringstates}, the states in the kernel of \eqref{eq:boundarycondop} are linear combinations of scattering states, namely 
\be
\ket{\boldsymbol \theta}^{s}=\sum_{\boldsymbol a}\Psi^{s}_{\boldsymbol a}(\boldsymbol \theta)\ket{\boldsymbol \theta}_{\boldsymbol a},\quad s=1,\ldots,2^N
\label{eq:linearcombination}
\ee
with 
\be
{\cal T}(\lambda|\boldsymbol \theta)^{\boldsymbol b}_{\boldsymbol a} \Psi^{s}_{\boldsymbol b}(\boldsymbol \theta)= \Lambda(\lambda|\boldsymbol \theta)^s\Psi^{s}_{\boldsymbol a}(\boldsymbol \theta),\quad s=1,\ldots,2^N.
\label{eq:eigensystem}
\ee
For each state \eqref{eq:linearcombination} with $s=1,\ldots,2^N,$ the condition of being in the kernel of the operators \eqref{eq:boundarycondop} is turned into a set of $N$ algebraic equations for the rapidities 
\be
\Lambda(\theta_j|\boldsymbol \theta)^s=e^{-i L M c \sinh \theta_j},\qquad j=1,\ldots,N,
\label{eq:BE}
\ee
known as ``Bethe equations''. Note that, as a consequence of \eqref{eq:SYB},
\be
\left[{\cal T}(\lambda|\boldsymbol \theta),{\cal T}(\mu|\boldsymbol \theta)\right]=0,\qquad \mu,\lambda\in\mathbb R\,,
\ee
so $\{{\cal T}(\theta_j|\boldsymbol \theta)\}_{j=1}^N$ are all simultaneously diagonalisable.  

The coefficients $\Psi^{s}_{\boldsymbol b}(\boldsymbol \theta)$ and the eigenvalues $\Lambda(\theta_j|\boldsymbol \theta)^s$ can be determined via an algebraic Bethe ansatz construction~\cite{BeSE14, FeTa11}. In this framework, instead of the index $s$, the states are conveniently parametrised by $1 \leq r \leq N$ auxiliary parameters ${\boldsymbol\lambda_r=(\lambda_1,\ldots,\lambda_r)}$, namely 
\be
\{\Psi^{s}_{\boldsymbol a}(\boldsymbol \theta)\}_{s=1}^{2^N}\to \{\Psi_{\boldsymbol a}(\boldsymbol \theta | \boldsymbol \lambda_r)\}_{r=1}^{N}\,.
\ee 
These parameters can be thought of as the rapidities of some fictitious particles called ``magnons'' that carry no energy and momentum but encode the topological charge structure of the multi-soliton states. Each magnon corresponds to a ``topological-charge-flip'' in the reference state which is taken to be the all-soliton state [i.e., ${\boldsymbol a=(+,+,\ldots,+)}$ in Eq. \eqref{eq:scatteringstate}]. Consequently, the topological charge of a state with $N$ physical particles and $r$ magnons is $Q=N-2r$.

The algebraic Bethe ansatz construction ensures that one can find a $\Psi_{\boldsymbol a}(\boldsymbol \theta | \boldsymbol \lambda_r)$ fulfilling \eqref{eq:eigensystem} if the magnonic rapidities satisfy 
\be
\label{eq:magnons}
\prod_{k=1}^{N}\!S_T(\lambda_{j}-\theta_{k})^{-1}\!=\!\prod_{k\neq j}^{r}\!\frac{S_T(\lambda_{k}-\lambda_{j})}{S_T(\lambda_{j}-\lambda_{k})} \,,\;\;\; j=1,\ldots r,
\ee
where $S_T(\theta)$ is defined in \eqref{eq:ST}. In this case the eigenvalue reads as 
\begin{align}
\Lambda(\lambda|\boldsymbol \theta| \boldsymbol \lambda_r)=& \left\{\prod_{k=1}^{N}S_T(\lambda-\theta_k)\prod_{k=1}^{r}S_T(\lambda-\lambda_k)^{-1}\right.\notag\\
&\left.+\prod_{k=1}^{r}S_T({\lambda}_k-\lambda)^{-1}\right\}\prod_{k=1}^{N}S_0(\lambda-\theta_k),
\label{eq:eigenvalue}
\end{align}
where $S_0(\theta)$ is introduced in \eqref{eq:S0}. Here we do not report an explicit expression of the coefficients as it is not needed. The interested reader can find it in Refs.~\cite{BeSE14, FeTa11}. 

Plugging \eqref{eq:eigenvalue} into the Bethe equations \eqref{eq:BE} we finally have
\be
\label{eq:quant}
\mathrm{e}^{i L Mc\sinh\theta_{j}}\prod_{k=1}^{r}S_{T}(\lambda_{k}-\theta_j)^{-1}\prod_{k=1}^{N}S_{0}(\theta_j-\theta_{k})=1
\ee
with $j=1,\ldots,N$. The coupled set of equations \eqref{eq:quant} and \eqref{eq:magnons} are known as ``nested'' Bethe Ansatz equations because they result from two subsequent applications of Bethe ansatz.

\subsection{Thermodynamic Bethe Ansatz}

In the large volume limit, the roots of the magnonic Bethe equations \eqref{eq:magnons} organise themselves in regular patterns in the complex plane called ``strings'', i.e., sets of complex roots with the same real part [called string centre] and imaginary parts differing by integer multiples of $i\pi$~\cite{Taka}. To classify all the possible string solutions it is useful to note that equations \eqref{eq:magnons} can be mapped to the Bethe Ansatz equations for the gapless XXZ spin-1/2 chain, and use the well known classification of the solutions of the latter~\cite{Taka}.

For generic values of $p$, strings of arbitrary length exist. In order to have a finite number of string species, we shall focus on the case where $p$ is an integer. In this case there are only $p$ different species of strings: ``positive parity" strings of length $1,2,\dots,p-1$ centred on the real axis and a ``negative parity" string of length 1 and imaginary part $p\pi/2$~\cite{Taka}. 

Once the string solutions have been identified, one takes the product of all equations referring to rapidities of the same string and writes a set of equations involving only the real string centres. In this way the equations are interpreted as the quantisation conditions in a system with $p+1$ species of particles, one physical and $p$ ``magnonic", interacting with a diagonal scattering matrix. This procedure is explicitly carried out in Appendix~\ref{app:bethetakahashi}. 

In the thermodynamic limit, the solutions of the quantisation conditions densely cover the real line and it is convenient to describe the states using their densities $\{\rho_k(\theta)\}_{k=0}^p$, known as ``root densities". Here we labelled the physical particles by ${k=0}$ and the magnons by ${k=1,\dots,p}$ with species $p$ being the negative parity string of length 1. These densities can be thought of as generalisations of the ``occupation numbers'' commonly used to characterise states in free systems. It is also useful to introduce the so-called ``hole densities'' $\rhoh_k(\theta)$ representing the densities of values of the rapidities allowed by the quantisation conditions but not populated in the specific state. As a consequence of the quantisation conditions, root and hole densities are connected by the following equations (see Appendix~\ref{app:TBA} for an explicit derivation)
\be
\label{eq:rhoeq}
\!\!\rho_k(\th) +\rhoh_k(\th)  = \delta_{k,0}\frac{Mc\cosh\th}{2\pi}-\nu_k\!\!\sum_{m=0}^p (a_{km}*\rho_m)(\th)\,,
\ee
known as thermodynamic Bethe--Takahashi equations. Here $\nu_k=1$ for ${1\le k\le p-1}$ and ${\nu_0=\nu_p=-1}$; the notation 
\be
(f*g)(\la)=\int \frac{\ud \mu}{2\pi} f(\la-\mu)g(\mu)
\ee
denotes the convolution, and the explicit expressions for the kernels $a_{nm}(\theta)$ are given in Appendix~\ref{app:TBA}. As customary in TBA, one can exploit a set of identities relating the kernels $a_{km}(\th)$ [see Appendix~\ref{app:TBA}] to cast the Bethe equations \eqref{eq:rhoeq} in a partially decoupled form, in which each species is coupled only to a small subset of other species. For $p>2$ the result reads [see Appendix~\ref{app:TBA} for $p=2$]
\bes
\label{eq:rhodec}
\begin{align}
\rhot_0(\th) &= \frac{M c \cosh\th}{2\pi} +s*\rhoh_1(\th)\,,\\
\rhot_1(\th) &= s*(\rho_0 +\rhoh_2+\delta_{p,3}\,\rho_3)(\th)\,,\\
\rhot_k(\th) &= s*(\rhoh_{k-1} +\rhoh_{k+1})(\th),\quad 1<k<p-2\,,\\
\rhot_{p-2}(\th) &= s*(\rhoh_{p-3} +\rhoh_{p-1}+\rho_p)(\th),\;\,\qquad  p>3\,,\label{rhop-2}\\
\rhot_{p-1}(\th) &=\rhot_p(\th)= s*\rhoh_{p-2}(\th)\,,\label{rhop-1p}
\end{align}
\esu
where we introduced the ``total" root densities
\be
\rhot_k(\theta) \equiv \rho_k(\theta)+\rhoh_k(\theta)\,,
\ee
and the function 
\be
s(\th)\equiv\frac1{\cosh\th}\,.
\ee
Note that the coupling pattern of the different species corresponds to the topology of the $D_{p+1}$ Dynkin diagram \cite{Zam91}.

The standard assumption of TBA is that, in the thermodynamic limit, the root densities $\{\rho_k(\th)\}_{k=0}^p$ of a given eigenstate of the Hamiltonian fully specify the expectation values of all local observables in that eigenstate. In other words, they give all the necessary information about the state. In general, however, the expression of the expectation value of local observables in terms of the root densities are unknown. An important exception are the densities of [local] conserved charges. Specifically, the expectation value of the conserved charge density $\hat o(t,x)$  in an eigenstate characterised by $\{\rho_k(\th)\}_{k=0}^p$ reads as 
\be
o =\sum_{k=0}^p  \int\ud\th \,\,o_{k}(\th) \rho_{k}(\th)\,,
\label{eq:chargeev}
\ee
where the functions $\{o_k(\th)\}_{k=0}^{p}$ are known as ``bare charges". For example, considering the expectation values of the densities of particles [$n$], energy [$e$], momentum [$p$], and topological charge [$q$] we have
\bes
\label{eq:bare}
\begin{align}
n_k(\th) &= \delta_{k,0}\,, & e_k(\th) &=  \delta_{k,0} M c^2 \cosh\th\,,\\
p_k(\th) &= \delta_{k,0}M c \sinh\th\,, & q_k(\th) &= 3 \delta_{k,0}-2\ell_k\,,
\end{align}
\esu
where $\ell_j$ is the length of the $j$-th string, namely $\ell_j=j$ for $1\le j<p$ and $\ell_0=\ell_p=1$. So,
\bes
\label{eq:NEQ}
\begin{align}
n &= \int\ud\th\,  \rho_0(\th)\,,\label{eq:N}\\
e &= \int\ud\th\,  M c^2 \cosh\th\, \rho_0(\th)\,,\label{eq:E}\\
p &= \int\ud\th\,  M c \sinh\th\, \rho_0(\th)\,,\label{eq:P}\\
q &= \int\ud\th\,  \rho_0(\th) - 2\sum_{m=1}^p  \ell_m\!\int\ud\th\,  \rho_m(\th)\,.\label{eq:Q}
\end{align}
\esu
More general examples of expectation values of local charges of the sine--Gordon model can be found in Ref.~\cite{VeCo17}. Note that a form like \eqref{eq:chargeev} holds for all expectation values of local conserved charge densities in any TBA solvable model. 

Together with the conserved charge densities, there is another class of local observables of which the expectation values are explicitly known in terms of the root densities. These are the ``currents'' of conserved charges, defined in terms of the charge densities through an operatorial continuity equation. More specifically, given a conserved charge $\hat O$ with density $\hat o(t,x)$, its current $\hat J_o(t,x)$ is defined by 
\be
\partial_t \hat o(t,x)+\partial_x \hat J_o(t,x) = 0 \,,
\ee
and ${\rm tr}[\hat J_o(t,x)]=0$. The expectation value of $\hat J_o(t,x)$ in a state described by  $\{\rho_k(\th)\}_{k=0}^p$ reads as~\cite{BCDF16, CaDY16} 
\be
{J}_o =\sum_{k=0}^p  \int\ud\th \,\,o_{k}(\th) \rho_{k}(\th)v_k^\text{dr}(\th)\,.
\label{eq:currentev}
\ee
Here $\{o_k(\th)\}_{k=0}^{p}$ are the same functions as those appearing in the expectation value of $\hat o(t,x)$, while $\{v_k^\text{dr}(\th)\}_{k=0}^p$ are the group velocities of excitations on the state specified by $\{\rho_k(\th)\}_{k=0}^p$. The latter are defined as 
\be
v_m^\text{dr}(\la) = \frac{ {e_m^\text{dr}}'(\la)}{{p_m^\text{dr}}'(\la)}\,,
\ee
where $e_m^\text{dr}(\la)$ and ${p_m^\text{dr}}(\la)$ are respectively the energy and the momentum of the elementary excitation consisting in the addition of a particle of species $m$ and rapidity $\lambda$. Note that, as a consequence of the interactions, adding or removing a particle changes the total energy and momentum of the state by a quantity which is different from its bare energy and momentum. In Bethe ansatz this is manifested in the change of the rapidities of all the particles in a given state when a particle is added or removed. Using the quantization conditions for the string centres, one can write down the following linear integral equations for  the velocities [cf. Appendix~\ref{app:TBA}]
\be
\label{eq:vdr}
\rhot_k(\th)v_k^\text{dr}(\th) = \frac{e'_k(\th)}{2\pi} -\nu_k \sum_{m=0}^p \left(a_{km}*\rho_m v^\text{dr}_m\right)(\th)\,.
\ee
The form \eqref{eq:currentev} is conjectured to apply to all TBA solvable models. Currently, however, an explicit proof is only available for relativistic integrable field theories with diagonal scattering \cite{CaDY16, VuYo19}. In the case of the XXZ spin-1/2 chain, although not proven, Eq.~\eqref{eq:currentev} has been thoroughly tested against DMRG simulations~\cite{BCDF16, PDCB17}.

\subsubsection{Root densities of the thermal state}

Let us consider the system in a thermal equilibrium state 
\be
\boldsymbol \rho=\frac{e^{-\beta (\hat H-\mu_1 \hat N-\mu_2 \hat Q)}}{{\rm tr}\left[e^{-\beta (\hat H-\mu_1 \hat N-\mu_2 \hat Q)}\right]}\,,
\label{eq:Gibbs}
\ee
where $\beta^{-1}=T$ is the temperature, and the two chemical potentials $\mu_1$ and $\mu_2$ control the densities of particles and of topological charge, corresponding to the operators $\hat{N}$ and $\hat{Q},$ respectively. The state \eqref{eq:Gibbs} can be represented microcanonically in terms of an eigenstate of the Hamiltonian, and thus, in the thermodynamic limit, it corresponds to a set of root densities $\{\rho_k(\th)\}_{k=0}^p$. 

Following Yang and Yang~\cite{YaYa69}, these root densities can be obtained by minimising  the grand potential density
\be
g=e-\mu_1 n -\mu_2 q -T s
\ee
with respect to $\{\rho_k(\th)\}_{k=0}^p.$ The densities of energy, particles, and topological charge are expressed in terms of the root densities in  Eqs.~\eqref{eq:NEQ}, while the entropy density $s$ assumes the so-called Yang--Yang form
\begin{align}
s =& \sum_{m=0}^p  \int\ud\th  \left[\rho_m(\th)\ln(1+\eta_m(\th))\right.\notag\\
&+\sum_{m=0}^p  \int\ud\th \left.\rhoh_m(\th)\ln(1+\eta^{-1}_m(\th))\right]\,,
\end{align}
where we introduced the functions  
\be
\eta_m(\th) = \frac{\rhoh_m(\th)}{\rho_m(\th)}\,.
\label{eq:etas}
\ee
The minimization procedure, with the constraints \eqref{eq:rhoeq}, leads to the following TBA equations 
\bes
\label{eq:TBA}
\begin{align}
\ln\eta_0(\th) = & \frac{M c^2 \cosh\th-\mu_1-\mu_2}T\nonumber\\
&+\sum_{m=0}^p \nu_m a_{m0}*\ln(1+\eta_m^{-1})(\th)\,, \\
\ln\eta_k(\th) = &\frac{2\mu_2}T\ell_k+\sum_{m=0}^p \nu_m a_{mk}*\ln(1+\eta_m^{-1})(\th)\,,
\end{align}
\esu
where  $k=1,\dots,p$. We stress that these equations can also be brought to a partially decoupled form with the same structure of \eqref{eq:rhodec} [see Appendix~\ref{app:TBA}]. Finally we note that using  \eqref{eq:TBA} and \eqref{eq:rhoeq} one can prove that
\begin{align}
 g=-T\int\frac{\ud\th}{2\pi} M\cosh\theta\ln(1+\eta_0^{-1})(\th)\,. 
\label{eq:identity}
\end{align}

The equations presented so far provide a complete description of the thermodynamics of the model at equilibrium. Furthermore, as we will see in the following, they also represent a key ingredient for the analysis of bipartitioning protocols and, more generally, for a macroscopic description of the sine--Gordon theory out of equilibrium.

\section{bipartitioning protocol}
\label{sec:bipartitioning}

As we have discussed in Sec.~\ref{sec:intro}, in this work we investigate the non-equilibrium dynamics of the sine--Gordon field theory following a bipartitioning protocol. This protocol is simple enough to allow for exact statements, but, on the other hand, it displays all the interesting features associated with breaking of translational invariance at finite energy density.

More specifically, we are interested in the quantum quench from the initial state
\begin{equation}
\label{eq:ini_state}
\boldsymbol  \rho_0= \boldsymbol  \rho^{\text{L}}_0 \otimes \boldsymbol  \rho^{\text{R}}_0\,,
\end{equation}
where
\be
\!\!\boldsymbol  \rho^{\alpha}_0=\frac{e^{- \beta_{\alpha} ({H}_{\alpha}-\mu_{1,\alpha}{N}_{\alpha}-\mu_{2,\alpha}{Q}_{\alpha})}}{{\rm tr}\left[e^{- \beta_{\alpha} ({H}_{\alpha}-\mu_{1,\alpha}{N}_{\alpha}-\mu_{2,\alpha}{Q}_{\alpha})}\right]}\,,\,\quad \alpha\in\{\rm L,R\}\,,
\label{eq:leftrightstates}
\ee
and operators with subscripts L and R are defined by restricting the integrals of their density respectively to $x<0$ and $x>0$. 

We will study this problem by means of two different methods: (i) a generalisation of Sachdev's semiclassical approach, reviewed in Sec.~\ref{sec:semiclassics}; (ii) Generalised Hydrodynamics, reviewed in Sec.~\ref{sec:GHD}. A systematic comparison of the two approaches is carried out in Secs.~\ref{sec:comparisonlim} and \ref{sec:comparisongen}.

\subsection{Semiclassical Approach}
\label{sec:semiclassics}

The semiclassical approach considered in this work concerns quantum systems with multiple species of long-lived quasiparticle excitations distinguished by some internal degree of freedom [different species]. The method was introduced in \cite{SaYo97,SaDa97} to describe low-temperature correlations in a transverse field Ising chain, and, then, applied for the study of several systems both in \cite{DaSa98,DaSa05,RZ:semiclassics} and out of \cite{RSMS09,RiIg11,Evan13,KoZa16, MoKZ17, KoMZ18, WMLKZ19, RiIg11, IgRi11, DiIR11, BlRI12} equilibrium. The main idea is that when the density of quasiparticles is low, they propagate along classical trajectories. On the contrary, the collisions among the particles are governed by quantum mechanics. In 1D collisions are inevitable, but at low enough densities it is sufficient to consider 2-particle scatterings only.

Different versions of the method differ in how accurately they treat the scattering. In this section we consider the simplest approximation, called ``universal limit", in which one assumes that all quasiparticles have negligibly small momenta [see Sec.~\ref{sec:comparisongen} for refinements]. For local interactions, the range of  the scattering potential is then much smaller than the de Broglie wavelength of the particles, so the potential can be substituted by a Dirac-delta and the scattering matrix becomes that of non-relativistic hard-core particles. At small momenta this scattering matrix becomes independent of the incoming momenta and fully reflective in the space of internal quantum numbers \cite{Sachdev_book}. The simple nature of the two-body S-matrix turns the problem into a fully-deterministic classical many-particle problem. The expectation values of observables are obtained by averaging over an ensemble of possible initial configurations. This classical formulation allowed  for analytic calculations in various models and physical setups \cite{RSMS09, SaYo97,SaDa97,DaSa98,DaSa05,RZ:semiclassics,RiIg11,Evan13,KoZa16,KoMZ18, RiIg11, IgRi11, DiIR11, BlRI12}. In particular, at large space- and time- scales, this problem can be treated by means of a hydrodynamic approach~\cite{Spohnbook}, where the state of the system is assumed to be locally stationary and completely fixed by the conserved quantities of the flow. 

Note that the above description directly applies to the case of the sine--Gordon field theory in the repulsive regime. In this case, as discussed in Sec.~\ref{sec:hamandS}, the relevant quasiparticles are solitons and antisolitons and the internal degree of freedom is the topological charge~\eqref{eq:tcharge}. Since the model is integrable, the quasiparticles are stable [infinitely long-lived] and the scattering is exactly factorised into sequences of 2-body processes for any density. As explicitly shown in Eqs.~\eqref{eq:Smatrix}, however, the 2-body scattering matrix generically depends non-trivially on the momentum. It takes the universal form~\eqref{eq:zeromomentum} only at small rapidities. This situation is realised, for instance, after a bipartitioning protocol, when the temperatures of the two leads \eqref{eq:leftrightstates} are low enough.

In the case of the bipartitioning protocol, one assumes that quasiparticles on the left (L) and right (R) leads are initially evenly distributed in space with densities $n_{\rm R/L}$, and their momenta and charges are drawn from the factorised distribution 
\be
f_{\alpha}(Q,p) = f_{\alpha}(p)g_{\alpha}(Q)\,,\qquad \alpha\in\{\rm L, R\},
\label{eq:initialdist}
 \ee
where $p\in\mathbb R$ is the momentum, ${Q\in\{+,-\}}$ is the internal charge, and we have 
\be
\int \!{\rm d}p\, f_{\alpha}(p)=\sum_{Q\in\{\pm\}}g_{\alpha}(Q)=1\,,\qquad \alpha\in\{\rm L, R\}.
\label{eq:initialdistnorm}
\ee 
We point out that the factorised form \eqref{eq:initialdist} is a necessary ingredient for the feasibility of the semiclassical calculation, however, the explicit form of $f_{\alpha}(Q)$ and $g_{\alpha}(Q)$ is not~\cite{KoMZ18}. Here we keep them generic and specify them in order to match the GHD solution. Moreover, in writing \eqref{eq:initialdist} and \eqref{eq:initialdistnorm} we specialised to the case where the internal charge takes only two distinct values, as it is the relevant one for us. The treatment, however, can be directly generalised to higher number of species.

Due to one dimensional kinematics, the collisions are fully elastic: momenta are always transmitted and topological charges are always reflected. This suggests to adopt a modified picture of the dynamics, in analogy with the treatment of one-dimensional hard rods~\cite{hardrods, DShardrods}. We split each quasiparticle in momentum and charge ``tracers" that have respectively fixed momentum and fixed internal charge. Initially, the momentum and charge tracers corresponding to the same particle move with the same velocity, but upon scattering momentum tracers are transmitted while charge tracers are reflected. Moreover, since the quasiparticles are point-like, tracers of momentum $p$ move along straight lines with slope 
\be
v(p)=\frac{{\rm d}\varepsilon(p)}{{\rm d}p} = \frac{p}{M}\,,
\ee
where we introduced the quasiparticle dispersion 
\be
\varepsilon(p)=M c^2 + \frac{p^2}{2M}\,,
\ee
and, since the universal limit applies at small momenta, we considered the non-relativistic form.

The dynamics can be translated in a hydrodynamic language [see, e.g., Ref.~\cite{Spohnbook}] by introducing $n(p,x,t)$, the density of tracers of momentum $p$ averaged over the distribution of initial configurations. The microscopic conditions on the scattering described above imply that $n(p,x,t)$ is locally conserved, namely  
\be
\partial_t n(p,x,t) + \partial_x J(p,x,t)=0 \qquad \forall p\,,
\label{eq:continuitynp}
\ee
where $J(p,x,t)$ is the current of tracers with momentum $p$. Since momentum tracers move along straight lines we simply have 
\be
J(p,x,t)= v(p) n(p,x,t)\,.
\label{eq:fixedpcurrent}
\ee
Plugging this into \eqref{eq:continuitynp} we find  
\be
n(p,x,t)= n_\text{L}(p) \Theta[v(p)-x/t]+ n_\text{R}(p)  \Theta[x/t-v(p)],
\ee
where $\Theta[x]$ is the Heaviside step function and we introduced $n_{\alpha}(p)$, the density of tracers of momentum $p$ in the lead ${\alpha=\{\rm L, R\}}$ averaged over the initial configurations. Using \eqref{eq:initialdist} we have   
\be
n_{\alpha}(p)= n_\alpha f_{\alpha}(p)\,.
\label{eq:Boltzmannnumber}
\ee
Going back to the quasiparticle formulation, $n(p,x,t)$ is the density of quasiparticles of momentum $p$, counting both those with positive and those with negative internal charge. This quantity can be used to compute the expectation values of observables to which the two species of particles contribute in the same way. For example, the total density of quasiparticles and the corresponding current read as  
\begin{align}
n(x,t) =& \int{\ud p}\, n(p,x,t)\,,\label{eq:npsc}\\
J_n(x,t) =& \int{\ud p} \,v(p)n(p,x,t)\,.
\label{eq:jpsc}
\end{align}
The energy density and current are given by analogous expressions with the integrands containing an extra factor of $\varepsilon(p)$. Note that these expressions depend solely  on the scaling variable 
\be
\zeta=x/t\,,
\ee
i.e., the ``ray'' in the $(x,t)$ plane. Accordingly, we use the notation 
\be
n(x,t)\mapsto n(\zeta)\,,\qquad J_n(x,t)\mapsto J_n(\zeta)\,.
\ee 

It is easy to see that $n(p,x,t)$ are not the only conserved quantities in the classical problem under exam. Indeed, the number of charge tracers of each given species is also conserved. By construction, the total density of charge tracers coincides with \eqref{eq:npsc}, but the difference between the number of charge tracers of the two species is an independent conserved quantity. This is nothing but the the topological charge. In hydrodynamic language we then have  
\be
\partial_t q(x,t) + \partial_x J_q(x,t)=0\,,
\label{eq:continuityq}
\ee
where $q(x,t)$ is the of topological charge density and $J_q(x,t)$ the corresponding current. As opposed to momentum tracers, however, the charge tracers scatter nontrivially, and the expression of $J_q(x,t)$ in terms of the conserved quantities is significantly more complicated than \eqref{eq:fixedpcurrent}. This expression is reported in Ref.~\cite{KoMZ18} and it is rather unwieldy, as it contains derivatives of the conserved quantities. In the scaling limit
\be
\lim_{{\rm sc}, \zeta}\equiv \lim_{\substack{x,t\to\infty \\ x/t=\zeta}}\,, 
\label{eq:scalinglimit}
\ee 
however, all derivatives can be neglected and we have
\be
\lim_{{\rm sc}, \zeta} J_q(x,t) =v(\zeta) q(\zeta)\,.
\ee
Here we introduced the ``fluid velocity"
\be
v(\zeta) \equiv \frac{J_n(\zeta)}{n(\zeta)}\,.
\label{eq:tcurrent}
\ee
Plugging \eqref{eq:tcurrent} in \eqref{eq:continuityq} we obtain the Euler equation for the topological charge. Its solution reads as   
\be
\lim_{{\rm sc}, \zeta} q(x,t)  =  n(\zeta) ( \tilde q_{\rm R} \Theta[\zeta-v(\zeta)] + \tilde q_{\rm L} \Theta[v(\zeta)-\zeta] )\,,
\label{eq:scaling_limit_sc}
\ee   
where we introduced the topological charge per particle in the lead $\alpha\in\{\rm L,R\}$
\be
\tilde q_{\alpha}=  \sum_{Q\in\{\pm\}} Q\, g_{\alpha}(Q)\,.
\ee
Interestingly, we see that both the density and the current of the topological charge feature a jump at the ray $v^*$, defined as the solution of 
\be
\zeta-v(\zeta)=0\,.
\label{eq:defvstar}
\ee
This jump, however, is not a physical shock: the subleading corrections to $J_q(x,t)$ show that it is a diffusively broadening front. The associated diffusion constant reads as~\cite{KoMZ18}
\be
\label{Dstar}
D^*=\frac{\Gamma(v^*)}{n(v^*)^2}\,,
\ee
where 
\be
\begin{split}
\Gamma(v^*)&=\int{\ud p}\, \Theta[v^*-v(p)]n_\text{R}(p)(v^*-v(p))\\
&=\int{\ud p}\, \Theta[v(p)-v^*]n_\text{L}(p)(v(p)-v^*)
\end{split}
\ee
is the flux of particles [from left to right and from right to left].

\begin{figure*}
	\centering
	\includegraphics[scale=0.8]{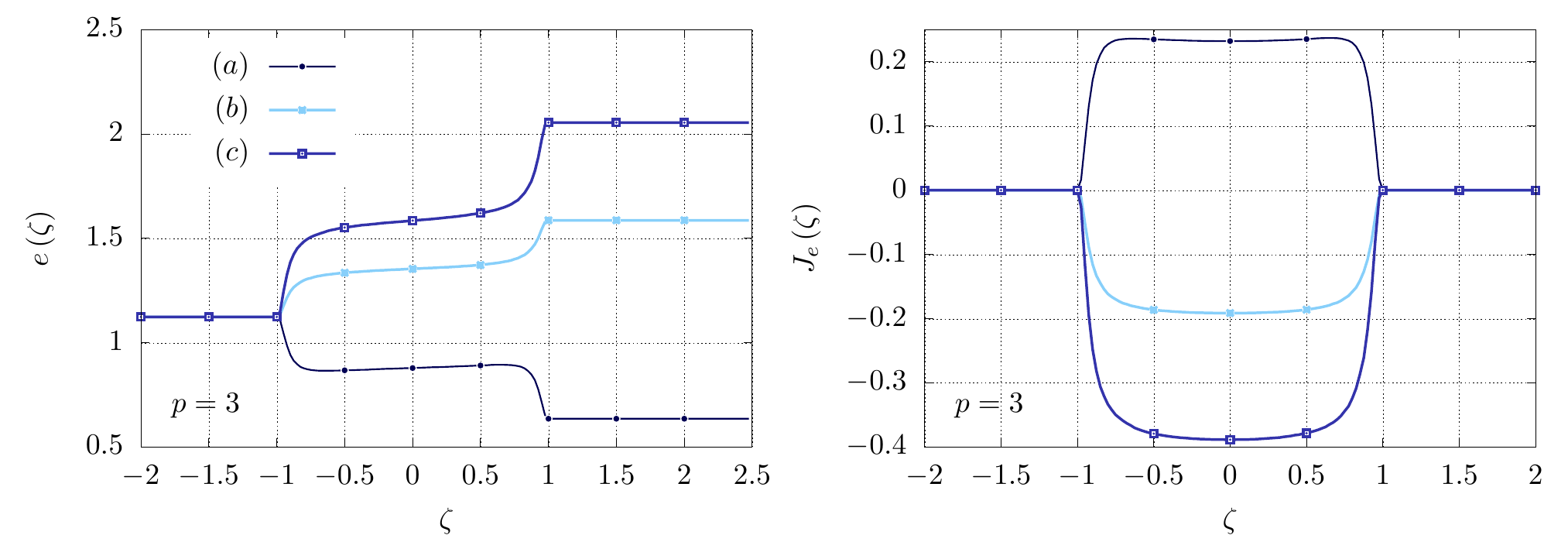}
	\caption{Profiles of energy density (left) and current (right) as a function of the velocity $\zeta=x/t$ at infinite times after a quench from the bipartite state. The left half is initially prepared in a state with $\beta_{\rm L}=\mu_{1,{\rm L}}=\mu_{2, \rm L}=1$, while the parameters corresponding to the right initial state are: $(a)$ $\beta_{\rm R}=2$, $\mu_{1,{\rm R}}=0.5$, $\mu_{2,{\rm R}}=1$;  $(b)$ $\beta_{\rm R}=1$, $\mu_{1,{\rm R}}=1$, $\mu_{2,{\rm R}}=1$; $(c)$ $\beta_{\rm R}=1$, $\mu_{1,{\rm R}}=0.5$, $\mu_{2,{\rm R}}=2$. All plots correspond to $p=3$ and $c=1$.}
	\label{fig:energy_profiles}
\end{figure*}

\subsection{Generalised Hydrodynamics}
\label{sec:GHD}

\begin{figure*}
	\centering
	\includegraphics[scale=0.8]{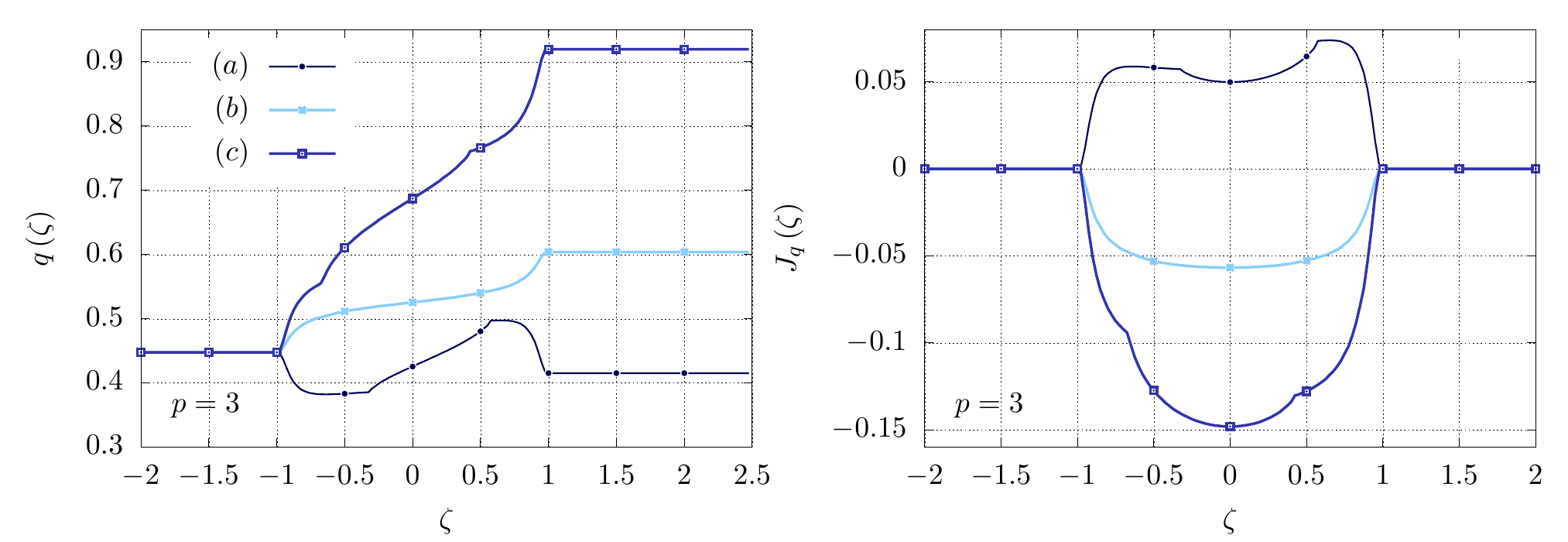}
	\caption{Profiles of topological charge (left) and topological current (right) as a function of the velocity $\zeta=x/t$ at infinite times after a quench from the bipartite state. The left half is initially prepared in a state with $\beta_{\rm L}=\mu_{1, \rm L}=\mu_{1, \rm L}=1$, while the parameters corresponding to the right initial state are: $(a)$ $\beta_{\rm R}=2$, $\mu_{1, \rm R}=0.5$, $\mu_{2, \rm R}=1$;  $(b)$ $\beta_{\rm R}=1$, $\mu_{1, \rm R}=1$, $\mu_{2, \rm R}=1$; $(c)$ $\beta_{\rm R}=1$, $\mu_{1, \rm R}=0.5$, $\mu_{2, \rm R}=2$. All plots correspond to $p=3$ and $c=1$. }
	\label{fig:charge_profiles}
\end{figure*}

The GHD treatment of the bipartitioning protocol relies on very different premises and analyses the problem fully quantum-mechanically. The starting point is the assumption that, at large times $t$ and distances $x$ from the junction, expectation values of local observables are described by a slowly varying stationary state, specified by an appropriate set of root densities $\{\rho_{n,x,t}(\th)\}_{n=0}^p$~\cite{BCDF16,CaDY16}. Accordingly, observables display non-trivial profiles connecting the values corresponding to the two reservoirs far away from the junction. Exploiting the conservation of the infinitely many charges of the integrable Hamiltonian, it is possible to write down a set of continuity equations connecting root densities at different points. These equations are the central result of the theory and read as~\cite{CaDY16, BCDF16} 
\be
\partial_t\rho_{n,x,t}(\th)+\partial_x(v^\text{dr}_{n,x,t}(\th) \rho_{n,x,t}(\th))=0\,.
\label{eq:continuity}
\ee
Here $v^\text{dr}_{n,x,t}(\lambda)$ is the dressed velocity of excitations on the state described by $\{\rho_{n,x,t}(\lambda)\}$. It is determined by solving the Bethe--Takahashi equations \eqref{eq:rhoeq} combined with \eqref{eq:vdr} in the state specified by $\rho_{n,x,t}(\th)$, namely  
\begin{align}
\label{eq:rhoray}
&\rhot_{n,x,t}(\th) = \frac{p_{n}'(\th)}{2\pi} -\nu_n\sum_{m=0}^p (a_{nm}*\rho_{m,x,t})(\th)\,,\\
\label{eq:vrhoray}
&v^\text{dr}_{n,x,t}(\th)\rhot_{n,x,t}(\th)  = \frac{e_{n}'(\th)}{2\pi} \notag\\
&\qquad\qquad\qquad-\nu_n\sum_{m=0}^p (a_{nm}*v^\text{dr}_{m,x,t}\rho_{m,x,t})(\th)\,.
\end{align}
As shown in \cite{BCDF16,CaDY16}, one can write down an implicit solution to Eq.~\eqref{eq:continuity} as follows. First, one argues that the rapidity distribution functions only depend on the ray $\zeta=x/t$ in the case of the bipartitioning protocol. Next, one can derive from \eqref{eq:continuity} a set of equations for the ``filling functions"  
\be
\vartheta_{n,\zeta}(\th)=\frac{\rho_{n,\zeta}(\th)}{\rho^{t}_{n,\zeta}(\th)}=\frac{1}{1+\eta_{n,\zeta}(\th)}\,,
\label{eq:filling_functions}
\ee
[cf. Eq.~\eqref{eq:etas}], which read
\be
(v^\text{dr}_{n,\zeta}(\th)-\zeta)\partial_\zeta\vartheta_{n,\zeta}(\th)=0\,.
\ee
These equations are immediately solved by 
\be
\vartheta_{n,\zeta}(\th) = \![\vartheta_{n,\rm L}(\th)-\vartheta_{n,\rm R}(\th)]\Theta[v^\text{dr}_{n,\zeta}(\th)-\zeta]+\vartheta_{n,\rm R}(\th).
\label{eq:implicitsolution}
\ee
Here $ \vartheta_{n,\rm L/R}(\th)$ are the filling functions of the states on the two sides of the junction.  Even though \eqref{eq:implicitsolution} represents an implicit solution to the continuity equation [as the dressed velocities also depend on the filling functions], it allows us to obtain an explicit numerical solution to high precision. Indeed, Eq.~\eqref{eq:implicitsolution} can be solved by iteration: as a first step, one starts with an initial ansatz for the filling functions $\vartheta_{n,\zeta}(\th)$; next one uses this ansatz to compute the dressed velocities; finally, plugging these into \eqref{eq:implicitsolution} one obtains a better approximation for the filling function \eqref{eq:filling_functions}. This procedure is seen to converge very quickly. After a solution for $\vartheta_{n,\zeta}(\th)$ is found, one can plug the latter into the Bethe equations to obtain a final result for the rapidity distributions functions $\rho_{n,\zeta}(\th)$. The knowledge of $\rho_{n,\zeta}(\lambda)$ allows us to compute the space-time profiles of local observables. In particular, charge densities and currents are found using \eqref{eq:chargeev} and \eqref{eq:currentev}. Some relevant results are reported in Figs.~\ref{fig:energy_profiles} and \ref{fig:charge_profiles}, displaying the profiles of density and current of energy and topological charge, respectively, computed for $p=3$.

The key observation connecting the GHD approach with the semiclassical one is that \eqref{eq:continuity} can be interpreted as the hydrodynamic equations for the conserved quantities $\{\rho_{n,x,t}(\th)\}_{n=0}^p$ of a gas of ${p+1}$ species of classical particles. Actually, this can be done in two different ways. First, as noted already in Ref.~\cite{BCDF16}, one can view $\{\rho_{n,x,t}(\th)\}_{n=0}^p$ as the conserved numbers [at fixed rapidity $\theta$] of some \emph{free} ``macroscopic" classical particles, moving with a space-time dependent velocity $v^\text{dr}_{n,x,t}(\th)$ that encode all the effects of the interactions. The second, more microscopic, interpretation is given in Ref.~\cite{DoYC18}. One views $\{\rho_{n,x,t}(\th)\}_{n=0}^p$ as the conserved numbers [at fixed rapidity $\theta$] of some \emph{interacting} classical particles moving with space-time independent ``bare" velocities
\be
v_{n}(\th)=\frac{e_{n}'(\th)}{p_{n}'(\th)}\,,
\ee
where bare energy and momentum are defined in Eqs.~\eqref{eq:bare}. For example, in the case of the Lieb-Liniger model, the analogues of \eqref{eq:continuity} are obtained as the classical hydrodynamic equations of a ``flea gas": an interacting classical gas where the particles, upon colliding, jump forward or backward by a distance that depends on their velocities~\cite{DoYC18}.

Even if both GHD and the semiclassical approach can be formulated in terms of classical hydrodynamic problems they are manifestly inequivalent. Indeed, they  generically give qualitatively different predictions. This is clearly demonstrated by the topological charge density and current profiles reported in Fig.~\ref{fig:charge_profiles}: as opposed to the semiclassical prediction they do not show jumps. Moreover, we note that the topological charge profile clearly shows points of non-analyticity for $|\zeta|<1$ in correspondence to the velocity of magnons. This feature has been observed before in the context of GHD, for example for XXZ Heisenberg chains~\cite{PDCB17}, and it is again absent in the semiclassical prediction. Shedding some light on the connection between the two approaches is the objective of the two following sections.

\section{From GHD to semiclassics: analytic treatment}
\label{sec:comparisonlim}

Here we start our systematic comparison by showing that GHD recovers exactly the semiclassical solution in two limiting cases. First, we consider a particular non-relativistic limit of the sine--Gordon theory where the soliton scattering becomes fully reflective. In this case the assumptions of the semiclassical approach are \emph{exactly fulfilled} and should be recovered by GHD: in Sec.~\ref{sec:sanity} we show analytically that this is indeed the case. In Sec.~\ref{sec:lowT}, instead, we study the low temperature limit of the GHD profiles, and show that their leading order is correctly described by the semiclassical theory.

\subsection{Non-Relativistic Limit}
\label{sec:sanity}

The fully reflective scattering regime is achieved by considering the following non-relativistic limit 
\be
c\rightarrow\infty\,, \qquad c\,\beta_{\rm sG}^2=\text{fixed}\,,
\label{eq:def_double_scaling}
\ee
where we also keep fixed the momentum of any given soliton, so that we must accordingly scale its rapidity
\be
k = M c \sinh \theta(k) \quad \Rightarrow \quad \theta(k)=\frac{k}{M c}+O\left(\frac{1}{c^2}\right)\,.
\label{eq:fixedmomentum}
\ee
In this limit, the renormalised coupling and the soliton mass are fixed, but solitons and antisolitons have vanishing rapidity, so that their scattering matrix takes the ``universal form"~\eqref{eq:zeromomentum}. Note that in this case we are making no assumption on the density of particles. 

Taking the non-relativistic limit of the Bethe equations, and shifting the magnonic rapidities
\be
\lambda_j \mapsto \lambda_j + i{\pi}/{2}\,,
\ee
we find 
\be
\!\!\!\!\!\left[\frac{\sinh\left[(i{\pi}/{2}-{\la_j})/p\right]}{\sinh\left[(i{\pi}/{2}+{\la_j})/p\right]}\right]^N \!\!\!\!\!\!=\!\!\prod_{k\neq j}^r \frac{\sinh\left[({\la_j}-{\la_k}-i{\pi})/{p}\right]}{\sinh\left[({\la_j}-{\la_k}+i{\pi})/{p}\right]},\label{magneq2}
\ee
for $j=1,\dots,r$, and
\be
e^{iLk_j}\prod_{k=1}^r \frac{\sinh\left(i{\pi}/{2p}-{\la_k}/p\right)}{\sinh\left(i{\pi}/{2p}+{\la_k}/p\right)}=(-1)^N
\label{bethe}
\ee
for $j=1,\dots,N$. We see that the first system coincides with the Bethe equations of the XXZ spin-1/2 chain, apart from a factor $(-1)^{N+1}$ on the left hand side. Since this factor does not modify the rapidity distribution functions, we expect the TBA equations describing the magnonic sector to coincide with those of the XXZ spin-1/2 chain. The second equation, instead, is a free twisted quantisation condition for the rapidities $k_j$ 
\be
\prod_{k=1}^r \frac{\sinh\left(i{\pi}/{2p}-{\la_k}/p\right)}{\sinh\left(i{\pi}/{2p}+{\la_k}/p\right)}=e^{i \Lambda}\,,
\ee
where $\Lambda$ is the total momentum in the magnonic chain. Since the twist does not affect the rapidity distribution, we expect the TBA equations describing the distribution of real momenta $k_j$ in the semiclassical limit to be free. 

Computing explicitly the non-relativistic limit of the Bethe--Takahashi equations \eqref{eq:rhoeq} (see Appendix~\ref{sec:appendix_double_scaling} for details), we find that our expectations are indeed confirmed. In particular we obtain 
\bes
\label{eq:scBT}
\begin{align}
&\tilde \rho^{t}_{0}(k) = \frac{1}{2\pi}\,,\label{eq:scrho}\\
&\rhot_{k}(\th) =\frac{n}{2\pi}\nu_k a_k(\th) -\nu_k \!\!\sum_{m=1}^p (a_{km}*\rho_{m})(\th)\,,\label{eq:scmag}
\end{align}
\esu
where $1\leq k\leq p$, $n$ is the density of particles of the first species [cf. Eq.~\eqref{eq:NEQ}], and we introduced the total root density of real momenta 
\be
\tilde \rho^{t}_0(k) =  \rho^{t}_0(\theta(k))\frac{d\theta(k)}{dk}=\frac{\rho^{t}_0(\theta(k))}{Mc\cosh\theta(k)}\,.
\label{eq:rhotk}
\ee
Note that \eqref{eq:scrho} is a Bethe--Takahashi equation for free particles while \eqref{eq:scmag} are that of the XXZ spin-1/2 with anisotropy $\Delta=\cos(\pi/p)$ [originating from a chain of $nL$ sites]. The only non trivial coupling between the two equations is given by the density $n=N/L$. Introducing the root density of real momenta
\be
\tilde \rho_0(k) =  \rho_0(\theta(k))\frac{d\theta(k)}{dk}=\frac{\rho_0(\theta(k))}{M c \cosh\theta(k)}\,,
\label{eq:rhok}
\ee
we can express the density as   
\be
n=\!\int\!\!{\rm d}k'\, \tilde \rho_0(k')\,.
\label{eq:density}
\ee

We stress that the main assumption in the derivation of \eqref{eq:scBT} is that, in the non-relativistic limit, the root density of real momenta has support on a finite region. This assumption is analogous to \eqref{eq:fixedmomentum} and selects the relevant states in the non-relativistic limit: those where the particles have finite momentum.  

Proceeding in the same way, we can also compute the non-relativistic limit of the equations \eqref{eq:vdr} that give access to the dressed velocities. The result reads as 
\bes
\begin{align}
\tilde \rho^{(t)}_0(k)v^{\rm dr}_0(k) &= \frac{k}{2\pi M}\,,\label{eq:velscrho}\\
\!\!\rhot_k(\lambda)v^{\rm dr}_k(\lambda) &= \frac{J_n}{2\pi}\nu_k a_k(\th) -\nu_k \!\!\sum_{m=1}^p (a_{km}*\rho_m v^{\rm dr}_m)(\th)\,,\label{eq:velscmag}
\end{align}
\esu
where $1\leq k\leq p$ and we introduced the particle current [cf. \eqref{eq:currentev}]
\be
J_n=\!\int\!\! {\rm d}k' \, \tilde \rho_0(k')v_0(k')\,.
\label{eq:Jdensity}
\ee
Eq.~\eqref{eq:velscrho} is again completely decoupled from the rest. Combined with \eqref{eq:scrho} it implies that, in the non-relativistic limit \eqref{eq:def_double_scaling}, the physical particles move with a state-independent velocity, which corresponds to the group velocity of free non-relativistic particles,
\be
v^{\rm dr}_0(k)=\frac{k}{M}\,.\label{eq:velocitiesNR1}
\ee
Looking more closely, however, we see that also Eqs. \eqref{eq:velscmag} are special. As usual, these equations have the same form as the corresponding Bethe--Takahashi equations [cf.~\eqref{eq:scmag}]. In the case at hand, however, a much stronger property holds. Indeed, the driving terms are \emph{proportional}. This means that 
\begin{align}
\rho^{(t)}_m(\lambda) \frac{J_n}{n},
\qquad\text{and}\qquad
\rho^{(t)}_m(\lambda)v^{\rm dr}_m(\lambda)
\end{align}
solve the same equations. Assuming that the solution is unique [this is necessary for the TBA treatment to be well-defined] we then have that the two sets coincide, namely 
\be
v^{\rm dr}_m =\frac{ \rho^{(t)}_m(\lambda)v_m(\lambda)}{ \rho^{(t)}_m(\lambda)}\!=\!\frac{J_n}{n}\!=\frac{\int\! {\rm d}k' \, \tilde \rho_0(k')v_0(k')}{\int\!{\rm d}k' \, \tilde \rho_0(k')}\equiv v\,.\label{eq:velocitiesNR2}
\ee
This equation is remarkable. It states that the velocities of all the magnons coincide for any state, and are given by the fluid velocity $v$ of the physical particles. This also means that the magnons are non-dispersive, i.e., $v_m$ does not depend on $\lambda$, which immediately implies the presence of jumps in the profiles of local observables.   

Combining equation \eqref{eq:velocitiesNR1} with the solution \eqref{eq:implicitsolution} we have 
\be
\!\!\tilde{\vartheta}_{0,\zeta}(k)\!=\!\!\left[\tilde{\vartheta}_{0,\rm L}(k)-\tilde{\vartheta}_{0,\rm R}(k)\right]\!\!\Theta[k-M\zeta]+\tilde{\vartheta}_{0,\rm R}(k)\,,\label{eq:nonrelsol1}
\ee
where $\tilde{\vartheta}_{0,\zeta}(k)$ is the filling function for real momenta, defined as $\tilde{\vartheta}_{0,\zeta}(k)={\vartheta}_{0,\zeta}(\lambda(k))$, and the subscript ${\alpha\in\{\rm L,R\}}$ denotes filling functions of the thermal states in the two leads. Using \eqref{eq:scrho} we then find 
\be
\tilde{\rho}_{0,\zeta}(k)\!=\!\left[\tilde{\rho}_{0,\rm L}(k)-\tilde{\rho}_{0,\rm R}(k)\right]\Theta[k-M\zeta]+ \tilde{\rho}_{0,\rm R}(k)\,, 
\label{eq:splitting_rho_0}
\ee
where $\tilde{\rho}_{0,\rm L/R}(k)=\tilde{\vartheta}_{0,\rm L/R}(k)/2\pi$. Plugging $\tilde{\rho}_{0,\zeta}(k)$ in \eqref{eq:velocitiesNR2} we find $v({\zeta})$, the fluid velocity at ray $\zeta$, which fixes the velocity of all magnons. Substituting it in \eqref{eq:implicitsolution} we then have  
\be
\!\!\vartheta_{k,\zeta}(\lambda) \!=\!\! \left[\vartheta_{k,\rm L}(\lambda) -\vartheta_{k,\rm R}(\lambda) \right]\!\Theta[v({\zeta})\!-\zeta]\!+\vartheta_{k,\rm R}(\lambda)\,.\label{eq:nonrelsol2}
\ee
Note that the solutions \eqref{eq:nonrelsol1} and \eqref{eq:nonrelsol2} are totally explicit: $\vartheta_{0,\zeta}(\th)$ is fixed by \eqref{eq:velocitiesNR1}, while all other $\vartheta_{k,\zeta}(\th)$ depend only on $\vartheta_{0,\zeta}(\th)$ [through $v({\zeta})$]. Plugging \eqref{eq:nonrelsol2} into \eqref{eq:scmag} we arrive at 
\begin{widetext} 
\begin{align}
\!\!\frac{\rhot_{k,\zeta}(\th)}{n(\zeta)} =&\frac{1}{2\pi}\nu_k a_k(\th)-\!\left[ \nu_k \!\!\sum_{m=1}^p \left(a_{km}*\vartheta_{m,\rm L}\frac{\rho^t_{m,\zeta}}{n(\zeta)}\right)(\th)\right]\!\Theta[v({\zeta})\!-\zeta]\!-\!\left[\nu_k \!\!\sum_{m=1}^p \left(a_{km}*\vartheta_{m,\rm R}\frac{\rho^t_{m,\zeta}}{n(\zeta)}\right)(\th)\right]\!\Theta[\zeta-v({\zeta})],\label{eq:intermediatesteprho}
\end{align}
where $n(\zeta)$ is obtained by plugging \eqref{eq:splitting_rho_0} in \eqref{eq:density}. Consider now $\zeta>v(\zeta)$. In this case ${\rhot_{k, \zeta}(\th)}/{n(\zeta)}$ fulfils the same equation as ${\rhot_{k,\rm R}(\th)}/{n_{\rm R}}$, where $n_{\rm R}$ and $\rhot_{k,\rm R}(\th)$ are respectively the particle density and the total root density of the $k$-th species of magnons in the thermal state on the right lead. Analogous considerations hold for $\zeta<v(\zeta)$, with ${\rhot_{k,\rm R}(\th)}/{n_{\rm R}}$ replaced by ${\rhot_{k,\rm L}(\th)}/{n_{\rm L}}$. Since the solution of \eqref{eq:intermediatesteprho} must be unique we have 
\begin{align}
&\!\!\!\frac{\rhot_{k,\zeta}(\lambda)}{n(\zeta)} \!=\!\! \left[\!\frac{\rhot_{k,\rm L}(\lambda)}{n_{\rm L}}\! -\!\frac{\rhot_{k,\rm R}(\lambda)}{n_{\rm R}} \right]\!\!\Theta[v({\zeta})\!-\zeta]\!+\! \frac{\rhot_{k,\rm R}(\lambda)}{n_{\rm R}}\,,
\label{eq:splitting_rho_k_0}
\end{align}
which implies 
\begin{align}
&\!\!\!\frac{\rho_{k,\zeta}(\lambda)}{n(\zeta)} \!=\!\! \left[\!\frac{\rho_{k,\rm L}(\lambda)}{n_{\rm L}}\! -\!\frac{\rho_{k,\rm R}(\lambda)}{n_{\rm R}} \right]\!\!\Theta[v({\zeta})\!-\zeta]\!+\! \frac{\rho_{k,\rm R}(\lambda)}{n_{\rm R}}\,.
\label{eq:splitting_rho_k}
\end{align}
At this point, the only missing step for a complete solution of the GHD problem is an expression of the ``boundary conditions'' $\{\tilde\rho_{0,\rm L/R}(p),\rho_{k,\rm L/R}(\th)\}$. These are obtained in Appendix~\ref{sec:appendix_double_scaling} by writing and solving the non-relativistic limit of the TBA equations \eqref{eq:TBA}. The solution reads as
\bes
\begin{align}
&\tilde \rho_{0,\alpha}(k)=\frac{1}{2\pi}\left[1+\frac{\exp[{\frac{k^2}{2MT_{\alpha}}-\frac{\tilde \mu_{1,\alpha}}{T_{\alpha}}}]}{2\cosh(\frac{\mu_{2,\alpha}}{T_{\alpha}})}\right]^{-1}\,,\label{eq:rhothermalfreefermions}\\
&\rho_{k,\alpha}(\lambda)={\rho}_{k}\!\left(\!\lambda,\frac{\mu_{2,\alpha}}{T_{\alpha}},n_{\alpha}\right)\,,\qquad\qquad \alpha\in\{\rm L,R\}\,,
\end{align}
\esu
where we introduced 
\bes
\label{eq:rhosT}
\begin{align}
&{\rho}_{k}(\lambda,h,n)=\frac{n \tanh(h)\sinh(h)}{4\pi\sinh[(k+1) h]} \left[\frac{ a_k(\lambda)}{\sinh[kh]} -\frac{ a_{k+2}(\lambda)}{\sinh[(k+2) h]} \right]\,, \qquad\qquad\qquad 1 \leq k \leq p-2\,,\\
&{\rho}_{p-1}(\lambda,h,n)=\frac{n \tanh[h]\left(\sinh[h]+e^{-p h}\sinh[(p-1)h]\right)}{4\pi\sinh[(p-1) h]\sinh[p h]}{a_{p-1}(\lambda)}\,,\\
&{\rho}_{p}(\lambda,h,n)=\frac{n \left(e^{-p h}\sinh[h]+\sinh[(p-1)h]\right)}{4\pi\cosh[h]\sinh[p h]}{a_{p-1}(\lambda)}\,.
\end{align}
\esu
Note that, to have a non-trivial result in the limit \eqref{eq:def_double_scaling}, one has to shift $\mu_1$ by the soliton rest energy $M c^2$, namely
\be
\limsc (\mu_1 - M c^2)=  \tilde \mu_1\,. \label{eq:sm_parameters_1}
\ee

Putting everything together we obtain the expectation value of a generic charge density $\hat{o}(x,t)$ [cf. \eqref{eq:chargeev}] and the associated current $\hat J_o(x,t)$ [cf. \eqref{eq:currentev}] 
\bes
\label{eq:finalresult}
\begin{alignat}{2}
o(\zeta) &= \quad\;\int {\ud k}\,  o_{0}(\theta(k)) &\tilde \rho_{0,\zeta}(k)+& \frac{n(\zeta)}{n_{\rm L}} \Theta[v({\zeta})-\zeta] \left(o_{\rm L}-\int\ud k \,\,o_{0}(\theta(k)) \tilde \rho_{0,{\rm L}}(k)\right) \notag\\
&&+&  \frac{n(\zeta)}{n_{\rm R}} \Theta[\zeta-v({\zeta})] \left(o_{\rm R} - \int\ud\lambda \,\,o_{0}(\theta(k)) \tilde \rho_{0,{\rm R}}(k)\right)\,,
\label{eq:finalresult1}\\
J_o(\zeta) &= \int {\ud k}\,  \frac{k}{M}o_{0}(\theta(k)) &\tilde \rho_{0,\zeta}(k)+& \frac{J_n(\zeta)}{n_{\rm L}} \Theta[v({\zeta})-\zeta] \left(o_{\rm L}-\int\ud k \,\,o_{0}(\theta(k)) \tilde \rho_{0,{\rm L}}(k)\right) \notag\\
&&+&  \frac{J_n(\zeta)}{n_{\rm R}} \Theta[\zeta-v({\zeta})] \left(o_{\rm R} - \int\ud\lambda \,\,o_{0}(\theta(k)) \tilde \rho_{0,{\rm R}}(k)\right)\\
&= \int {\ud k}\,  \frac{k}{M} o_{0}(\theta(k)) &\tilde \rho_{0,\zeta}(k) +& v(\zeta) \left(o(\zeta)-\int {\ud k}\,  o_{0}(\theta(k)) \tilde \rho_{0,\zeta}(k)\right)\,,\notag
\label{eq:finalresult2}
\end{alignat}
\esu
where $o_{\alpha}$ is the expectation value of the charge in the lead $\alpha\in\{\rm L, R\}$. 
\end{widetext}

Equations \eqref{eq:finalresult} simplify when the charge is only sensitive to the physical particles, i.e., when ${o_{j}(k)=0}$ for all ${j=1,\ldots,p}$. In this case we obtain the non-interacting result 
\bes
\label{eq:freefermionic}
\begin{align}
o(\zeta) =& \int {\ud k}\,  o_{0}(\theta(k)) \tilde \rho_{0,\zeta}(k)\,,\\
J_o(\zeta) =&  \int {\ud k}\,  \frac{k}{M} o_{0}(\theta(k)) \tilde \rho_{0,\zeta}(k)\,. 
\end{align}
\esu
Another simplification occurs when the charge has $o_0(\theta)=1$, as, for instance, the topological charge. In this case we have 
\begin{align}
\!\!\!\frac{J_o(\zeta)}{J_n(\zeta) } =\frac{o(\zeta)}{n(\zeta)} = \Theta[v({\zeta})-\zeta]\frac{o_{\rm L}}{n_{\rm L}}+  \Theta[\zeta-v({\zeta})] \frac{o_{\rm R}}{n_{\rm R}}\,.
\label{eq:double_scaling_tcharge}
\end{align}
In particular, for the topological charge the boundary conditions are exactly evaluated to [see Appendix~\ref{sec:appendix_double_scaling}]   
\be
\frac{q_{\alpha}}{n_{\alpha}}=\tanh(\mu_{2,{\alpha}}  \beta_{\alpha})\,,\qquad\qquad \alpha\in\{\rm L,R\}\,. 
\label{eq:tchargeboundaries}
\ee
Note that \eqref{eq:freefermionic} and \eqref{eq:double_scaling_tcharge} [with the boundary conditions \eqref{eq:tchargeboundaries}] coincide with the semiclassical expressions \eqref{eq:npsc} and \eqref{eq:scaling_limit_sc} if one chooses
\bes
\label{eq:identifications}
\begin{align}
n_{\alpha}&=\int\!\!{\rm d}k'\, \tilde \rho_{0,\alpha}(k')\,,\\
f_{\alpha}(p)&=\frac{\tilde \rho_{0,\alpha}(p)}{n_\alpha}\,,\\
g_{\alpha}(Q)&=\frac{e^{\mu_{2,\alpha}Q}}{2\cosh \mu_{2,\alpha}}\,,\qquad\qquad \alpha\in\{\rm L,R\}\,,
\end{align}
\esu
where $\tilde \rho_{0,\alpha}(k')$ is defined in Eq.~\eqref{eq:rhothermalfreefermions}. This completes the derivation of the semiclassical predictions from the non-relativistic limit of the GHD equations.

 \begin{figure*}
	\centering
	\includegraphics[scale=0.82]{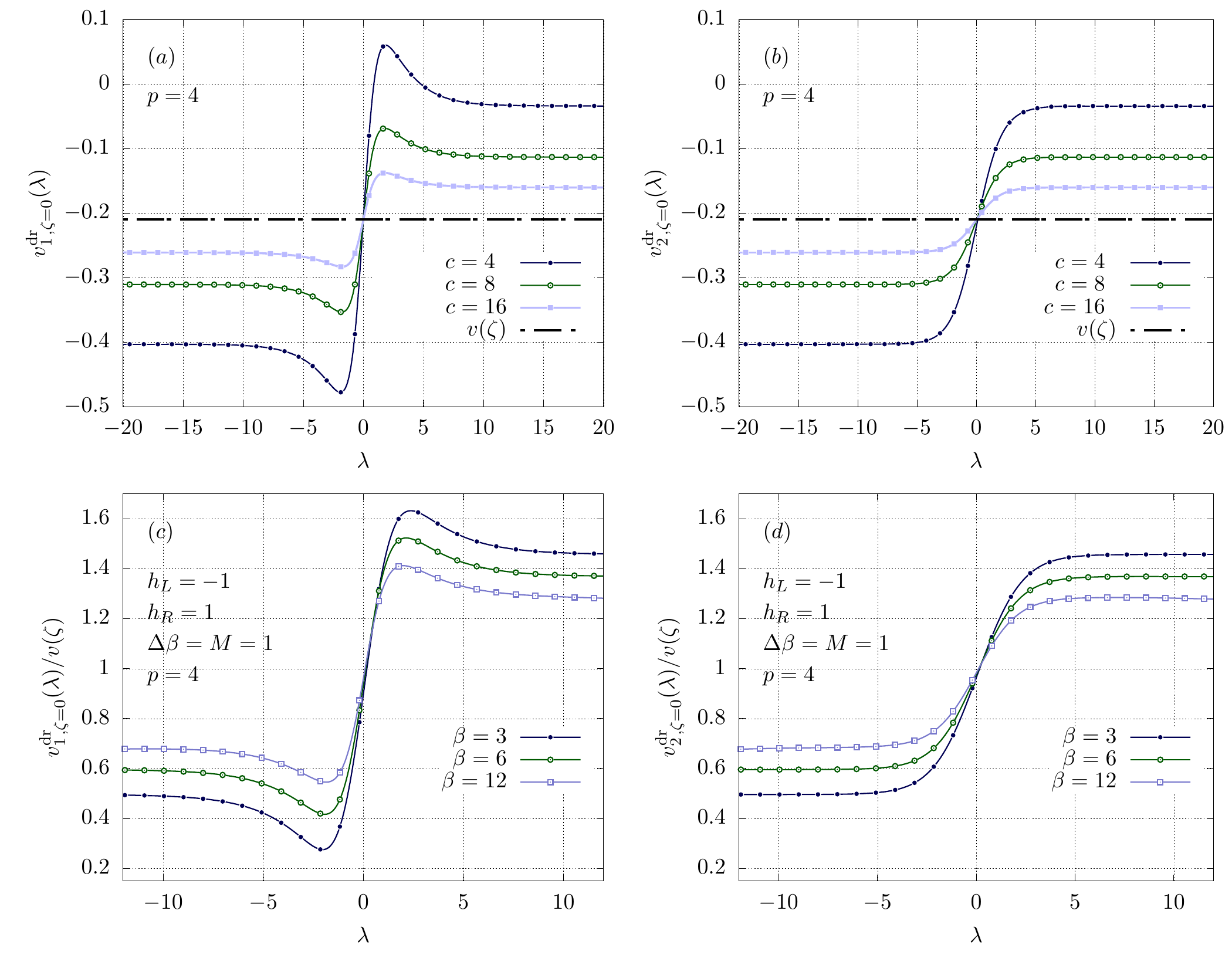}
	\caption{Dressed velocities $v^{\rm dr}_{j,\zeta}(\lambda)$ for the first magnons in the NESS. Subfigures $(a)$, $(b)$ show the behavior approaching the non-relativistic limit Eq.~\eqref{eq:def_double_scaling} by plotting $v^{\rm dr}_{j,\zeta}$ for increasing values of the speed of light $c$ while keeping $c\,\beta_{\rm sG}^2$ fixed. The initial bipartite thermal state corresponds to $M$=1, $\tilde{\mu}_{1,L}=1$, $\tilde{\mu}_{1,R}=2$, $\mu_{2,L}=0.25$, $\mu_{2,R}=0.5$, $\beta_L=1.2$ and $\beta_{R}=2.4$. The curves are clearly converging to the dashed line as the non-relativistic limit  is approached, indicating the value of $v(\zeta)$ defined in Eq.~\eqref{eq:velocitiesNR2}. Subfigures $(c)$ and $(d)$ correspond to the small temperature limit defined by Eqs.~\eqref{eq:def_lowT_1}--\eqref{eq:def_h2}. The parameters of the quench are displayed on the figure (with $\mu_{1,L}=\mu_{1,R}=0$). Since in this case $v(\zeta)$ vanishes as $\beta\to\infty$, the plots display the ratios $v^{\rm dr}_{1,\zeta}(\lambda)/v(\zeta)$ which are seen to converge to $1$.}
	\label{fig:velocities}
\end{figure*}

\subsection{Low Temperature Regime}
\label{sec:lowT}

In this section we study small temperatures: the second regime in which we expect to recover the semiclassical prediction from the GHD equations. In particular, we set 
\begin{align}
&\beta_{\rm L}=\beta\,,\label{eq:def_lowT_1}\\
&\beta_{\rm R}=\beta+\Delta \beta\,,\label{eq:def_lowT_2}
\end{align}
and consider an asymptotic expansion of the GHD solution for large $\beta$ while keeping $\Delta \beta$ fixed. Note that this is the natural setup to study low-temperature bipartitioning protocols in gapped systems. For instance, the analogous setup was studied in \cite{BePC18} for gapped XXZ Heisenberg chains. In our case we also fix 
\bes
\label{eq:def_h}
\begin{align}
&\beta_{\rm L}\mu_{2, \rm L}=h_{\rm L}\,,\label{eq:def_h1}\\
&\beta_{\rm R}\mu_{2, \rm R}=h_{\rm R}\,,\label{eq:def_h2}
\end{align}
\esu
while we assume that  $\mu_{1,\rm L}$ and $\mu_{1,\rm R}$ are strictly smaller than $Mc^2$ so that the system is gapped.

We point out that the limit studied here is conceptually very different from the one considered in the previous section. Previously the whole time-evolving Hamiltonian was rescaled, while here we are simply restricting to a special class of initial states. As we now show, however, the analytic treatment turns out to be very similar.  A simple way to see this is to consider the low-temperature limit of Eqs.~\eqref{eq:rhoray} and \eqref{eq:vrhoray}. As shown in Appendix~\ref{sec:appendix_small_T_limit}, the leading order contribution reads as 
\begin{align}
&\tilde\rho^t_0(k)=\frac{1}{2\pi}\,,\\
&\tilde\rho^t_0(k)v^{\rm dr}_{0}(\theta)=\frac{1}{2\pi} \frac{c k}{\sqrt{M^2c^2+k^2}}\,,
\end{align}
\begin{align}
&\rhot_{k}(\lambda) = \nu_n (a_{n}*\rho_{0,\zeta})(\lambda)-\nu_k \sum_{m=1}^p \left(a_{km}*\rho_{m}\right)(\lambda),\\
&\rhot_{n,\zeta}(\lambda)  v^{\rm dr}_{n,\zeta}(\lambda)  =  \nu_n (a_{n}*\rho_{0,\zeta} v^{\rm dr}_{0,\zeta})(\lambda)\notag\\
&\quad\qquad\qquad\qquad\quad - \nu_n \sum_{m=1}^p (a_{n m}*\rho_{m,\zeta} v^{\rm dr}_{m,\zeta})(\lambda)\,,
\end{align}
where we again used the root density of real momenta [cf.~\eqref{eq:rhok}]. The first two equations immediately give 
\be
v^{\rm dr}_{0}(k)=\frac{c k}{\sqrt{M^2c^2+k^2}}\,,
\label{eq:lowT_v1}
\ee
namely, the physical particles are once again moving as free particles, this time, however, with a relativistic dispersion. Moreover, expanding the driving terms of the last two equations for small temperatures we have [see Appendix~\ref{sec:appendix_small_T_limit}] 
\bes
\begin{align}
(a_{n}*\rho_{0,\zeta})(\lambda)&=\frac{1}{2\pi} a_{n}(\lambda) n(\zeta)+O(\beta^{-1})\,,\label{eq:driving1mt}\\
(a_{n}*\rho_{0,\zeta}v^{\rm dr}_{0,\zeta})(\lambda)&= \frac{1}{2\pi} a_{n}(\lambda) J_n(\zeta)+O(\beta^{-1})\,,\label{eq:driving2mt}
\end{align}
\esu
where $n(\zeta)$ and $J_n(\zeta)$ are again the physical particle density and current at ray $\zeta$. Crucially, we see that at the leading order \eqref{eq:driving1mt} and \eqref{eq:driving2mt} are again proportional, and we can repeat the argument of Sec.~\ref{sec:sanity} to find  
\be
v^{\rm dr}_{n,\zeta}(\lambda)=v(\zeta)+O(\beta^{-1})\,.
\label{eq:lowT_v2}
\ee
Here $v(\zeta)= J_n(\zeta)/n(\zeta)$ is again the fluid velocity of the physical particles. 

Due to the validity of Eqs. \eqref{eq:lowT_v1} and \eqref{eq:lowT_v2}, the analysis of the previous section directly applies also to the current case. In particular, the root densities are again given by the equations \eqref{eq:splitting_rho_0} and \eqref{eq:splitting_rho_k}. The only difference is that in this case physical particles have a relativistic dispersion. This means that the velocity $v_0(k)$ is given by \eqref{eq:lowT_v1} and $\tilde\rho_{0,\rm L/R}(k)$ read as
\begin{align}
\!\!\!\!\tilde \rho_{0,\alpha}(k)\!=\!\! \frac{1}{2\pi}\! \left[\!1\!+\!\frac{\exp[{\sqrt{({M^2c^4+c^2 k^2}-\mu_{1,\alpha})}/{T_{\alpha}}}]}{2\cosh(h_{\alpha})}\!\right]^{-1}\!\!\!\!\!\!\!\!.
\end{align}
At the leading order in $\beta$, however, relativistic and non-relativistic dispersions produce the same predictions for the profiles of local observables. This can be seen by noting that both $\tilde \rho_{0,\rm L}(k)$ and $\tilde \rho_{0,{\rm R}}(k)$ are peaked around $k=0$ with a width that scales with $\beta^{-1}$. So that
\be
\frac{\int {\rm d}k\, f(k)(\tilde \rho_{0,\zeta}(k)-\tilde \rho^{\rm NR}_{0,\zeta}(k))}{\int {\rm d}k\, f(k)\tilde \rho_{0,\zeta}(k)}=O(1/\beta)\,,
\ee
where $\tilde \rho^{\rm NR}_{0,\zeta}(k)$ is the root density constructed with a non-relativistic dispersion and $f(k)$ is a generic smooth function. In summary, at the leading order in $1/\beta$, our results are again in agreement with the semiclassical predictions \eqref{eq:npsc} and \eqref{eq:scaling_limit_sc}  supplemented with the identifications \eqref{eq:identifications}.

We stress that the asymptotic expansion considered here is valid only if 
\be
\lim_{\beta\to\infty}Mc^2-\mu_{1,\alpha}>0\,, \qquad\forall \alpha\in\{\rm L,R\}\,,
\ee
meaning that the density of real particles must be exponentially small in $\beta$ in order for the semiclassical predictions to apply. This is  in contrast to the limit studied in the previous subsection, where the density of real particles is arbitrary.

 \begin{figure}
	\centering
	\includegraphics[scale=0.78]{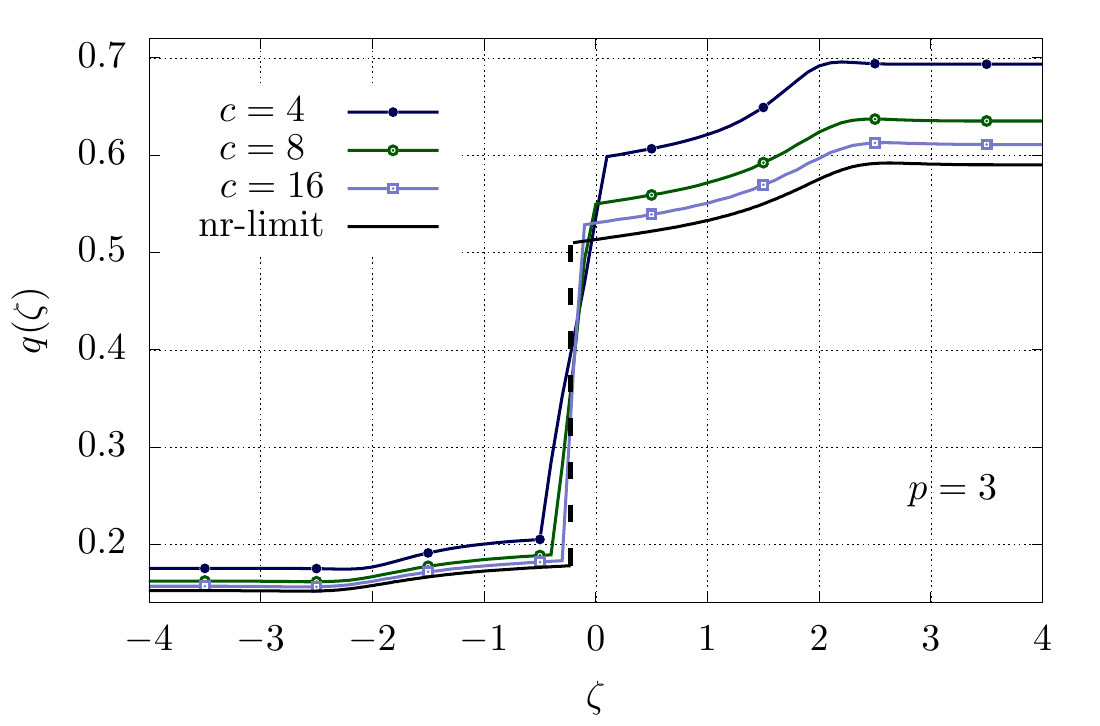}
	\caption{Comparison between the semiclassical result \eqref{eq:scaling_limit_sc} and the numerical evaluation of the GHD equations in the non-relativistic limit \eqref{eq:def_double_scaling}. The curves correspond to $M$=1, $\tilde{\mu}_{1,L}=1$, $\tilde{\mu}_{1,R}=2$, $\mu_{2,L}=0.25$, $\mu_{2,R}=0.5$, $\beta_L=1.2$ and $\beta_{R}=2.4$. Different curves correspond to increasing values of the speed of light $c$. The solid black line is the semiclassical result \eqref{eq:double_scaling_tcharge}, which displays a discontinuity at $\zeta^\ast\simeq -0.225$.}
	\label{fig:tCharge_double_scaling}
\end{figure}

\section{Finite energy scales: breakdown of the semiclassical approach}
\label{sec:comparisongen}

\begin{figure*}
	\includegraphics[scale=0.85]{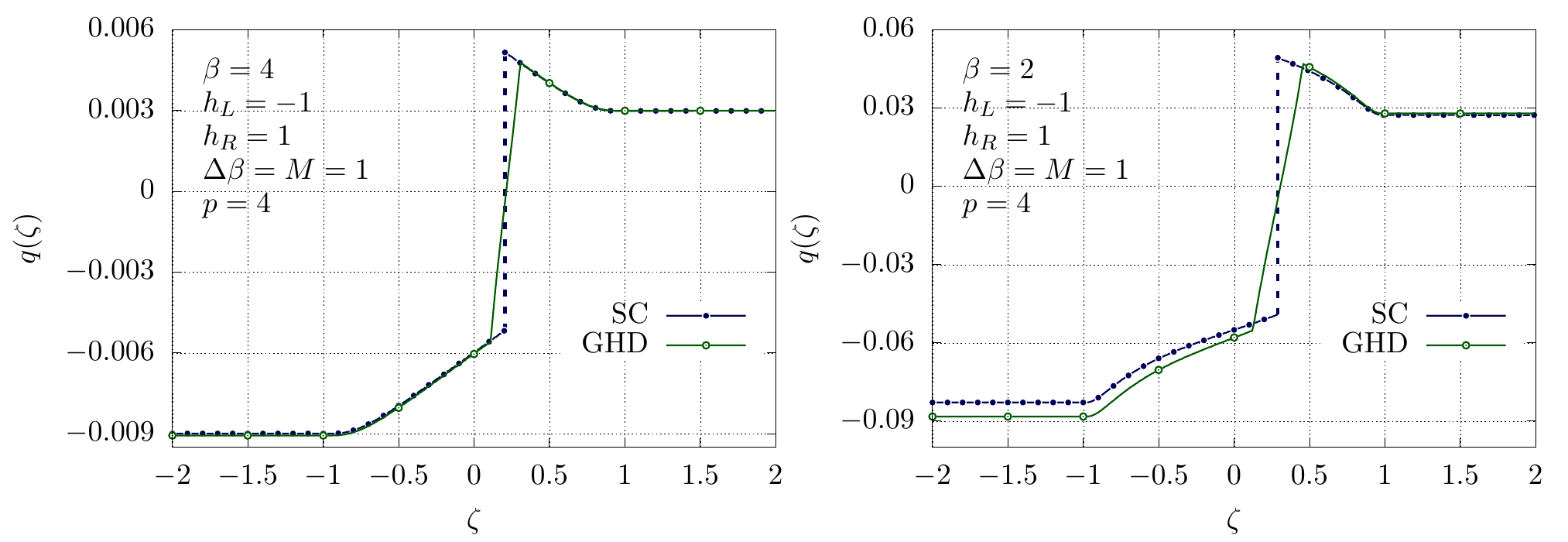}
	\caption{Comparison between the semiclassical result \eqref{eq:scaling_limit_sc} and the numerical evaluation of the GHD equations in the low temperature limit. The plots correspond to a bipartitioning protocol in the small temperature limit defined by Eqs.~\eqref{eq:def_lowT_1}--\eqref{eq:def_h2}. The parameters of the quench are displayed on the figure (with $\mu_{1,L}=\mu_{1,R}=0$ and $c=1$).}
	\label{fig:tcharge_lowT}
\end{figure*}

In this section we compare GHD and the semiclassical approach in the case of finite energy scales and densities of particles. As we have seen in the previous section, the two methods provide the same predictions in the non-relativistic limit \eqref{eq:def_double_scaling} and in the low-temperature regime~\eqref{eq:def_lowT_1}--\eqref{eq:def_h}. A common feature of both of these cases is that the velocities of  all magnons approach the same constant value, cf. Eqs. \eqref{eq:lowT_v1} and \eqref{eq:lowT_v2}. This is the crucial property that enables one to recover the semiclassical prediction. Indeed, as discussed in Sec.~\ref{sec:semiclassics}, in the semiclassical approach ``charge tracers" -- carrying information on the topological charge -- move with a \emph{single} velocity $v(\zeta)$ [cf. \eqref{eq:tcurrent}]. This feature is explicitly demonstrated in Fig~\ref{fig:velocities}, where we report the dressed velocities $v^{\rm dr}_{1,\zeta=0}(\lambda)$ and $v^{\rm dr}_{2,\zeta=0}(\lambda)$ of the first two magnons for various bipartitioning protocols. In particular, in the panels $(a)$ and $(b)$ we show data for the non-relativistic limit, where the chemical potential is correctly rescaled as in \eqref{eq:sm_parameters_1}. As $c$ increases, the curves are clearly converging to the constant value $v(\zeta)$, defined in Eq.~\eqref{eq:velocitiesNR2}.  In the panel $(c)$ and $(d)$, instead, we report data for small temperatures. The velocities $v^{\rm dr}_{j,\zeta}(\lambda)$ are now rescaled with $v({\zeta})$ [defined in \eqref{eq:lowT_v2}], which is vanishing as $\beta\to\infty$. In particular, since $v({\zeta})=O(\beta^{-1/2})$ [see Appendix~\ref{sec:appendix_small_T_limit}], we have $v^{\rm dr}_{k,\zeta}(\lambda)/v({\zeta})=1+O(\beta^{-1/2})$. This explains the large finite-$\beta$ corrections observed in the plot. 

Importantly, however, Fig~\ref{fig:velocities} also shows that when the value of $c$ and $\beta$ are finite, the velocities $v_{n,\zeta}(\lambda)$ are not constant and we expect this to cause a deviation of the profiles from the semiclassical prediction. Our numerical results show that this expectation is confirmed. Consider, for example, Fig.~\ref{fig:tCharge_double_scaling}, where we report the comparison between the semiclassical prediction and GHD numerical results obtained at finite values of $c$. We see that as $c$ increases, the profiles of the topological charge approach the semiclassical prediction, as they should. For finite values of $c$, however,  there is no abrupt jump [depicted in the figure by dashed lines], and the profiles remain continuous.

A completely analogous scenario is observed for low temperatures. This is demonstrated in Fig.~\ref{fig:tcharge_lowT}, where we report a comparison between GHD numerical results and the semiclassical predictions. Note that in this case the magnitude of the topological charge decreases as the temperature is lowered. As for the non-relativistic limit we see that, outside of the light-cone $|\zeta|<1$, the semiclassical predictions provide very good quantitative approximation to the correct result. However, at any finite temperature the profiles are always continuous, in contrast to the semiclassical result.

These numerical observations can be explained by noting that the classical many-particle problem associated with GHD has more ``conserved modes" than that emerging from Sachdev's semiclassical approach. In particular, in GHD the magnonic rapidities -- carrying information on the topological charge -- can ``decouple" from the real momenta, and spread ballistically with a spectrum of different velocities. This phenomenon can be interpreted as a sort of spin-charge separation, and it is in stark opposition with what happens in the semiclassical approach, where the topological charge degrees of freedom are ``dragged" by the fluid of physical particles.

In turn, this observation suggests that the main weakness of the semiclassical approach is not in the assumption of classical trajectories nor in that of low momenta, but is in the ``choice of quasiparticles", or, in other words, in the choice of the classical problem to which the original one is mapped. In principle, instead of the ``bare" particles [solitons and antisolitons in our case], one would need to use the ``Bethe ansatz'' particles, namely physical particles and magnons. These are indeed the only ones allowing for a diagonalisation of the interaction and a separation of ``charge" [particle number] and ``spin" [topological charge].

\begin{figure*}
	\centering
	\includegraphics[scale=0.84]{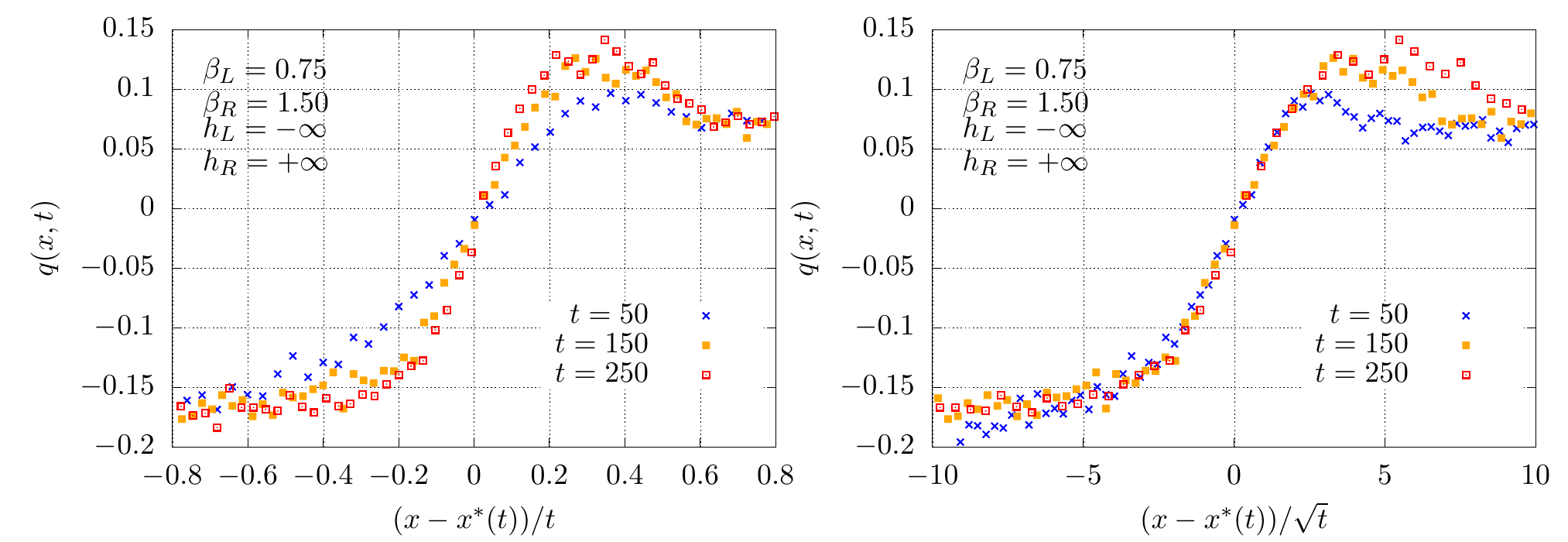}
	\caption{Hybrid ``semi-semiclassical'' profiles of the topological charge at finite times indicated in the plots after the quench from a bipartite state in the reference frame moving at the velocity $v^*,$ rescaled ballistically (left panel) and diffusively (right panel). The left half is initially prepared in a state with $\beta_{\rm L}=0.75,h_{\rm L}=-\infty$, while the parameters corresponding to the right initial state are: $\beta_{\rm R}=1.5,h_{{\rm R}}=\infty,$ so the two halves are completely ``polarized'' with opposite sign ($\mu_{1, \rm L}=\mu_{1,\rm R}=0$). The diffusive rescaling gives a good collapse of the finite time profiles near $x^*(t)=v^*t.$}
	\label{fig:hybridSC}
\end{figure*}

That being said, it is interesting to wonder whether some existing refinement of Sachdev's semiclassical method can overcome this weakness without changing the quasiparticle content. Here we considered what is arguably the most refined of all: the recently developed ``semi-semiclassical'' approach~\cite{MoKZ17}. In short, this is a hybrid method that describes the scattering of internal degrees of freedom fully quantum mechanically while maintaining the classical trajectories. This is achieved by building an effective ``spin chain" from the consecutive quantum numbers [topological charges] of the quasiparticles and evolving it by the exact two-body S-matrix whenever there is a collision of the semiclassical trajectories using a matrix product state algorithm [e.g. TEBD]. Note that, due to the full quantum mechanical treatment of the scattering, this method cannot be interpreted as a classical many-particle problem. Indeed, quantum superpositions are allowed in the internal space and possible quantum interference effects are also accounted for.

In this approach, charge degrees of freedom are still attached to the physical particles, but since there are finite amplitudes of transition and reflection, we cannot a priori exclude that it reproduces the GHD results. Simulations at finite times, however, indicate that even though the shape of the topological charge profile changes quantitatively, the broadening of the jump is still diffusive. This is demonstrated in Fig. \ref{fig:hybridSC} where the topological charge profiles, plotted at different times, are rescaled around the jump suggested by a ballistic and a diffusive broadening. The latter yields a better collapse of the profiles suggesting that the broadening remains diffusive even in the hybrid semiclassical approach. This sub-ballistic behaviour leads in the hydrodynamic scaling limit \eqref{eq:scalinglimit} to a discontinuous the profile, which is in contrast with the GHD results. For a definite conclusion, however, a more systematic study for larger times aimed at excluding a possible crossover from early diffusive to a late time ballistic behaviour would be necessary.

\section{Conclusions}
\label{sec:conclusions}

In this work we studied the dynamics of integrable quantum field theories with internal degrees of freedom in inhomogeneous settings by comparing the predictions of an out-of-equilibrium generalisation of Sachdev's semiclassical approach~\cite{SaYo97, SaDa97, KoMZ18}, with the exact result obtained via GHD~\cite{BCDF16, CaDY16}. Specifically, we focused on bipartitioning protocols in the sine--Gordon field theory. We identified two regimes where the semiclassical picture becomes exact: low temperatures and a particular non-relativistic limit. We also found that, away from these regimes, the semiclassical approach leads to incorrect qualitative predictions on the transport of topological charge. In particular, we found that at finite energies the topological charge always spreads ballistically, contrary to the semiclassical predictions. Finally, we showed that these discrepancies seem to remain in the improved ``hybrid" semiclassical approach of Ref.~\cite{MoKZ17}, where the evolution of internal degrees of freedom are described quantum-mechanically using DMRG simulations.

The limitations of the semiclassical picture are not to be ascribed to its classical nature, but rather to the choice of the quasiparticles which always correspond to the ``bare'' excitations of the field theory. Indeed, as shown in \cite{DoYC18}, the GHD equations can always be thought of as hydrodynamic equations of a classical many-particle system. In this case, however, the classical particles correspond to the ``Bethe ansatz" excitations that diagonalise the scattering. It is then interesting to wonder whether an improved semiclassical approach could be devised which considers such excitations as the quasiparticles.

\section*{Acknowledgments}

The authors thank J\'er\^ome Dubail, Benjamin Doyon, Olalla Castro-Alvaredo for useful discussions and, in particular, Pa\c{s}cu Moca for providing them with the numerical data for Fig.~\ref{fig:hybridSC}. B.~B. acknowledges support from the ERC under the Advanced grant number 694544 (OMNES) and from the Slovenian Research Agency (ARRS) under the grants P1-0402 and N1-0055. L.~P. acknowledges support from the Alexander von Humboldt foundation and from the Deutsche Forschungsgemeinschaft under Germany's Excellence Strategy -- EXC-2111 -- 390814868. M.~K. was supported by the Hungarian National Research, Development and Innovation Office (NKFIH) through Grant Nos. SNN118028. M.~K. was also supported by a ``Pr\'emium''  postdoctoral grant of the Hungarian Academy of Sciences. B.~B. thanks the Budapest University of Technology and Economics for hospitality. Part of this work has been carried out during the workshop ``Quantum Paths''  at  the  Erwin  Schr\"odinger  International Institute (ESI) in Vienna.


\onecolumngrid
\appendix
\section{Derivation of the Bethe--Takahashi equations}
\label{app:bethetakahashi}

In this appendix we provide further details on the derivation of the Bethe--Takahashi equations for the string centres.

We begin by recalling that, for generic values of $p,$ strings of arbitrary length exist. In order to have a finite number of string species, we shall focus on the case where $p$ is an integer. In this case there are only $p$ different string species: positive parity strings of length $1,2,\dots,p-1$ and a negative parity 1-string \cite{Taka}. 

Following the same steps as in the XXZ model, the magnonic equations \eqref{magneq2} can be rewritten in terms of the strings. We achieve this by multiplying the equations for the $\la_j$ belonging to a given string and on the right hand side we group the factors in the product over $\la_k$ according to the strings. Many factors cancel and the result can be written in compact form in terms of the real string centers. We introduce the notations
\be
g(\la,k)\equiv \frac{\sinh[\frac1p(\la-k\,i\frac\pi2)]}{\sinh[\frac1p(\la+k\,i\frac\pi2)]}\,,\qquad g^-(\la,k) = g\left(\la+ip\frac\pi2,k\right)= \frac{\cosh[\frac1p(\la-k\,i\frac\pi2)]}{\cosh[\frac1p(\la+k\,i\frac\pi2)]}\,.
\ee
For the product on the left hand side we use the identity 
\be
\prod_{\la_j\in n-\text{string at }\la} g(\la_j,1) = g(\la, n)\,,
\ee
while on the right hand side we encounter factors like
\begin{multline}
G_{n,m}(\la-\mu)\equiv\prod_{\la_j\in n-\text{string at }\la} \;\;\prod_{\mu_k\in m-\text{string at }\mu}g(\la_j-\mu_k,2) \\
= g(\la-\mu, |n-m|)^{1-\delta_{nm}}g(\la-\mu, |n-m|+2)^2\dots g(\la-\mu, n+m-2)^2g(\la-\mu, n+m)  \qquad 1\le n,m<p \,.
\end{multline}
We label the negative parity 1-string as the $p$-th string species for which

\bes
\begin{align}
G_{p,m}(\la-\mu)&=\prod_{\mu_k\in m-\text{string at }\mu}g(\la+ip\pi/2-\mu_k,2) =g^-(\la-\mu, m-1)g^-(\la-\mu, m+1)\qquad 1\le m<p\,,\\
G_{m,p}(\mu-\la)&=\prod_{\mu_k\in m-\text{string at }\mu}g(\mu_k-\la-ip\pi/2,2) =g^-(\mu-\la, m-1)g^-(\mu-\la, m+1)\qquad 1\le m<p\,,\\
G_{p,p}(\la-\mu)&=G_{1,1}(\la-\mu)=g(\la-\mu,2)\,.
\end{align}
\esu

Using these identities we find
\bes
\label{gGeqs}
\begin{align}
(-1)^{N \ell_j} \prod_{k=1}^N g(\la^{(j)}_\alpha-\th_k,n_j) & =  (-1)^{\ell_j}\prod_{k=1}^p \prod_{\beta=1}^{M_k} G_{j,k}(\la_\alpha^{(j)}-\la_\beta^{(k)})\qquad 1\le j<p\,,\\
(-1)^{N} \prod_{k=1}^N g^-(\la^{(p)}_\alpha-\th_k,1)  &=  -\prod_{k=1}^p \prod_{\beta=1}^{M_k} G_{p,k}(\la_\alpha^{(p)}-\la_\beta^{(k)})\,,\\
e^{iML\sinh\th_\alpha} \prod_{k=1}^{p-1} \prod_{\beta=1}^{M_k} g(\th_\alpha-\la_\beta^{(k)},k)^{-1}\;&\prod_{\beta=1}^{M_p}g^-(\th_\alpha-\la_\beta^{(p)},1)^{-1}
\prod_{\beta=1}^{N}S_{0}(\th_\alpha-\th_\beta)=(-1)^r\,,
\end{align}
\esu
where $\la^{(j)}_\alpha\in\mathbb{R}$ is the centre of the $\alpha$-th string of type $j$ [$\lambda^{(0)}_\alpha=\th_\alpha$], $M_k$ is the number of $k$-strings [$M_0=N$], and $\ell_j$ is the length of the $j$-string with $\ell_j=j$ for $1\le j<p$ and $\ell_0=\ell_p=1.$ 
The sign $(-1)^{\ell_j}$ on the right hand side comes from the fact that the unrestricted product now includes also the scattering of the roots with themselves and this gives a minus sign for each root in a string.  

Taking the logarithm of Eqs. \eqref{gGeqs} we obtain
\bes
\label{logeqs}
\begin{align}
\sum_{k=1}^N -i\log g(\la^{(j)}_\alpha-\th_k,n_j) & =  2\pi I^{(j)}_\alpha +\sum_{k=1}^p \sum_{\beta=1}^{M_k} -i\log G_{j,k}(\la_\alpha^{(j)}-\la_\beta^{(k)})\qquad 1\le j<p\,,\\
\sum_{k=1}^N -i\log g^-(\la^{(p)}_\alpha-\th_k,1)  &=  2\pi I^{(p)}_\alpha+\sum_{k=1}^p \sum_{\beta=1}^{M_k} -i\log G_{p,k}(\la_\alpha^{(p)}-\la_\beta^{(k)})\,,\\
ML\sinh\th_\alpha - \sum_{k=1}^{p-1} \sum_{\beta=1}^{M_k} -i\log g(\th_\alpha-\la_\beta^{(k)},k) -
&\sum_{\beta=1}^{M_p} -i\log g^{-}(\th_\alpha-\la_\beta^{(p)},1) +\sum_{\beta=1}^N -i\log S_0(\th_\alpha-\th_\beta) = 2\pi I^{(0)}_\alpha\,,
\end{align}
\esu
where $p_j(\la)=\delta_{j,0}\,m\sinh\la,$  $\Theta_{j,k}(\la)=-i\log G_{j,k}(\la),$ and $I^{(j)}_\alpha$ are integers or half-integers. The building blocks of the scattering phase shifts are 
\bes
\begin{align}
-i\log [-g(\la,k)] &= 2\arctan\left[\frac{\tanh\left(\frac\la{p}\right)}{\tan\left(\frac{k\pi}{2p}\right)}\right]\,,\\
-i\log g^-(\la,k)&= -2\arctan\left[\tanh\left(\frac\la{p}\right)\tan\left(\frac{k\pi}{2p}\right)\right]\,.
\end{align}
\esu
Whenever these phase shifts are at $k=p$ we should take $0$ for them.

\section{Details on the thermodynamic description of the sine--Gordon model}
\label{app:TBA}

In this appendix we derive in detail the thermodynamic limit of the Bethe equations. We start by recalling that, in this limit, the rapidities of the kinks as well as those of the magnons become dense, such that $L \rho_j(\la)\ud \la$ gives the number of roots for centres of $j$-strings. We have
\bes
\begin{align}
I_\alpha^{(j)} &= L\int^{\la_\alpha^{(j)}} \ud \la\,\rhot_j(\la)\qquad\qquad 0\le j<p\,,\\
I_\alpha^{(p)} &= L\int_{\la_\alpha^{(p)}} \ud \la\,\rhot_p(\la)\,,
\end{align}
\esu
where $\rhot_j(\la)=\rho_j(\la)+\rhoh_j(\la)$ is the total density of states (we label the physical particles by $j=0$), and we took into account that $ML\sinh\th$ and $-i\log g(\la,j)$ are monotonically increasing for $0\le j<p$ while $-i\log g^-(\la,1)$ is monotonically decreasing. Taking the derivative of the equations we get
\bes
\begin{align}
\sum_{k=1}^N a_{n_j}(\la^{(j)}_\alpha-\th_k) & =  2\pi L\rhot_j(\la_\alpha) +\sum_{k=1}^p \sum_{\beta=1}^{M_k} a_{j,k}(\la_\alpha^{(j)}-\la_\beta^{(k)})\qquad 1\le j<p\,,\\
\sum_{k=1}^N a_1^-(\la^{(p)}_\alpha-\th_k)  &=  -2\pi  L\rhot_p(\la_\alpha)+\sum_{k=1}^p \sum_{\beta=1}^{M_k} a_{p,k}(\la_\alpha^{(p)}-\la_\beta^{(k)})\,,\\
ML\cosh\th_\alpha - \sum_{k=1}^{p-1} \sum_{\beta=1}^{M_k} a_k(\th_\alpha-\la_\beta^{(k)}) -
&\sum_{\beta=1}^{M_p} a_1^{-}(\th_\alpha-\la_\beta^{(p)}) +\sum_{\beta=1}^N \varphi_0(\th_\alpha-\th_\beta) = 2\pi I^{(0)}_\alpha\,,
\end{align}
\esu
where
\be
\varphi_0(\th)=-i\frac{\ud}{\ud\th} \log S_0(\th) = \frac12 \int_{-\infty}^{\infty}\ud\w\frac{\sinh\frac{\pi\w}2(p-1)}{\sinh\frac{\pi\w p}2  \,\cosh\frac{\pi \w}2} e^{i\w\theta}\,,
\ee
and
\bes
\label{eq:app_anmdef}
\begin{align}
a_{nm}(\la)&=(1-\delta_{nm})a_{|n-m|}(\la)+2a_{|n-m|+2}(\la)+\dots +2a_{n+m-2}(\la)+a_{n+m}(\la)  \qquad 1\le n,m<p \,,\\
a_{pm}(\la)&=a_{mp}(\la)= (1-\delta_{m1})a^-_{m-1}(\la)+ (1-\delta_{m,p-1})a^-_{m+1}(\la)\qquad 1\le m<p\,,\label{eq:app_apm}\\
a_{pp}(\la)&=a_{11}(\la)=a_2(\la)
\end{align}
\esu
with
\bes
\begin{align}
a_n(\la) &= \frac2p \frac{\sin\left(\frac{n\pi}p\right)}{\cosh\left(\frac{2\la}p\right)-\cos\left(\frac{n\pi}p\right)}\,,\\
a^-_n(\la) &= -\frac2p \frac{\sin\left(\frac{n\pi}p\right)}{\cosh\left(\frac{2\la}p\right)+\cos\left(\frac{n\pi}p\right)}\,.
\end{align}
\esu
As discussed above, we should take $a^-_p(\la)\equiv0$ which explains the $1-\delta_{m,p-1}$ factor in Eq. \eqref{eq:app_apm}.

We can write all the equations in a unified form defining
\be
p_n(\theta) = \delta_{n,0}\,m\sinh\th
\ee 
and
\bes
\begin{align}
a_{m0}&=a_{0m}=-a_m\qquad\qquad\qquad\qquad\qquad 1\le m\le p-1\,,\\
a_{p0}&=a_{0p}=-a_1^-\,,\qquad a_{00} = \varphi_0\,.
\end{align}
\esu
With these notations we find 
\be
\label{p'L}
p_j'(\la^{(j)}_\alpha)L+\sum_{k=0}^p \sum_{\beta=1}^{M_k} a_{jk}(\la_\alpha^{(j)}-\la_\beta^{(k)}) =  -\nu_j 2\pi L\, \rhot_j (\la^{(j)}_\alpha)\,,
\ee
where $\nu_j=1$ for $1\le j\le p-1$ and $\nu_0=\nu_p=-1.$
Eqs. \eqref{logeqs} can also be written in a similar form, 
\be
\label{discrBE}
p_j(\la^{(j)}_\alpha)L+\sum_{k=0}^p \sum_{\beta=1}^{M_k} \Phi_{jk}(\la_\alpha^{(j)}-\la_\beta^{(k)}) =  (2\delta_{j,0}-1) 2\pi I^{(j)}_\alpha\,,
\ee
where
\bes
\begin{align}
\Phi_{jk} (\la)&= -i\log G_{jk}(\la)\qquad\qquad j,k=1,\dots,p\,,\\
\Phi_{0k}(\la)&=i\log g(\la,k)\,,\qquad  \Phi_{0p}(\la)=i\log g^-(\la,1)\,,\qquad \Phi_{00}(\th) = -i\log S_0(\th)\,.
\end{align}
\esu

In the thermodynamic limit, the sums over string centers can be written as integrals, which leads to
\be
\label{rhoeq1}
-\nu_n\rhot_n(\th) =\frac{p'_n(\th)}{2\pi}+\sum_{m=0}^p (a_{nm}*\rho_m)(\th)
\ee
or
\be
\rhot_n(\th) = \delta_{n,0}\frac{M\cosh\th}{2\pi}-\nu_n\sum_{m=0}^p (a_{nm}*\rho_m)(\th)\,.
\ee
Note that from $a_{p-1,m}=-a_{p,m}$ it follows that
\be
\label{rho_p-1rho_p}
\rhot_{p-1}(\th) = \rhot_p(\th)\,.
\ee

These equations can be brought to a partially decoupled form by exploiting functional identities linking the kernels $a_{nm}(\theta)$ with different indices \cite{Taka}. In particular, the following identities hold for $0\le m\le p$ (for $p>2$): 
%
\bes
\label{sids}
\begin{align}
a_{1m}&=s* a_{2m}+s(\delta_{2,m}-\delta_{0,m}-\delta_{m,p}\delta_{p,3})\,,\label{sa2m}\\
 a_{nm}&=s*( a_{n-1,m}+ a_{n+1,m})+s (\delta_{n-1,m}+\delta_{n+1,m}-\delta_{n,p-2}\delta_{m,p})\qquad\qquad 1<n<p-1\,,\label{sanm}\\
 a_{p-1,m}=- a_{p,m}&=s* a_{p-2,m}+s\,\delta_{p-2,m} 
\,,\label{sap-2}\\
 a_{0m} &= -s*a_{1m}-s\,\delta_{1,m}\,,\label{sa1m}
\end{align}
\esu
%
where 
\be
s(\theta) = \frac1{\cosh(\theta)}\,.
\ee
Note that 
\be
a_{00}=\varphi_0 = -s*a_{10}= s*a_1
\ee
which extends the standard XXZ identities and links the sine--Gordon kink scattering phase shift to those of the magnonic strings.

For $p=2$
\be
\label{asp2}
a_1=-a_{01}=a_{01^-}=-a_{1^-} = s\,,\qquad a_{11}=a_{11^-}=0\,,\qquad s*a_1 = a_{00}=\varphi_0\,,
\ee
where
\be
\varphi_0(\theta) = \frac14\int_{-\infty}^\infty \ud\w \frac1{\cosh^2\left(\frac{\pi \w}2\right)} e^{i\w\theta} = \frac{\theta}{\pi\sinh\theta}\,.
\ee

Following the steps in \cite{Taka} one arrives at the decoupled equations \eqref{eq:rhodec}. For the sake of completeness, we also give the equations for $p=2:$
\bes
\begin{align}
\rhot_0 &= \frac{M\cosh\th}{2\pi} + s*(\rhoh_1+\rho_2)\,,\\
\rhot_1 =\rhot_2&= s*\rho_0 \,.
\end{align}
\esu
The density, energy, and topological charge density in the thermodynamic limit are given by
\be
\label{NEQ}
n= \int\ud\th  \rho_0(\th)\,,\quad e= \int\ud\th  M\cosh\th \rho_0(\th)\,,\quad
q = \int\ud\th  \rho_0(\th) - 2\sum_{m=1}^p  \ell_m\int\ud\th  \rho_m(\th)\,,
\ee
and, as there are at most as many magnons as solitons ($r\le N),$
\be
\int\ud\th  \rho_0(\th) - \sum_{m=1}^p  \ell_m\int\ud\th  \rho_m(\th)\ge0\,.
\ee

To fully specify the thermodynamic state a relation is needed between the density of roots and the density of holes. This is encoded in the filling functions 
\be
\eta_m(\th) = \frac{\rhoh_m(\th)}{\rho_m(\th)}\,.
\ee
In thermal equilibrium, they can be obtained by minimizing the generalized free energy density
\be
g= e - Ts - \mu_1 n -\mu_2 q\,.
\ee
The standard procedure leads to the TBA equations \eqref{eq:TBA}. These equations can also be written in a partially decoupled form. For $p>2$ the resulting equations are 
\bes
\label{TBA2}
\begin{align}
\ln\eta_0&=\frac{M\cosh\th}T -s*\ln(1+\eta_1)\,,\label{eta0}\\
\ln\eta_1&=s*\ln\left[(1+\eta_0^{-1})(1+\eta_2)\right]+\delta_{p,3}s*\ln(1+\eta_p^{-1})\,,\label{eta1}\\
\ln\eta_n&=s*\ln\left[(1+\eta_{n-1})(1+\eta_{n+1})\right]\qquad\qquad 1<n<p-2\,,\\
\ln\eta_{p-2}&=s*\ln\left[(1+\eta_{p-3})(1+\eta_{p-1})(1+\eta_p^{-1})\right]\,,\\
\ln\eta_{p-1}&=p\frac\mu{T}+s*\ln(1+\eta_{p-2})\,,\label{etap-1p}\\
\ln\eta_p&=p\frac{\mu}{T}-s*\ln(1+\eta_{p-2})\,.\label{etap}
\end{align}
\esu
For $p=2$ they read
\bes
\begin{align}
\ln\eta_0&=\frac{M\cosh\th}T -s*\ln[(1+\eta_1)(1+\eta_2^{-1})]\,,\\
\ln\eta_1&=\frac{2\mu}T+s*\ln(1+\eta_0^{-1})\,,\\
\ln\eta_2&=\frac{2\mu}T-s*\ln(1+\eta_0^{-1})=\frac{4\mu}T-\ln\eta_1\,.
\end{align}
\esu
Finally, we consider the elementary excitations of over a state. For this it is useful to return to the Bethe--Takahashi equations \eqref{discrBE} in finite volume. Excitations correspond to changing a finite number of the integers $\{I_\alpha^{(j)}\}.$ Since the equations are coupled, all rapidities will be shifted. Even though these shifts are $O(L^{-1}),$ the number of rapidities is $O(L)$ leading to an $O(1)$ contribution to the  momentum, energy as well as to all the other conserved charges. So the bare charges of the excitation get ``dressed up'' by the presence of the sea of other particles.
Following the standard treatment using the shift function \cite{Korepin_book}, the dressed charges are given by
\be
\pm q^\text{dr}_m(\la) =  q_m(\la) +  \sum_{k=0}^p \int\frac{\ud\la'}{2\pi} h_k(\la')\Phi_{km}(\la'-\la)\,,
\label{eq:qdr}
\ee
where $h_k(\la)$ is the solution of the integral equation
\be
\label{heq}
h_k(\la) = q'_k(\la)\nu_k\vartheta_k(\la) -  \sum_{n=0}^p  \int\frac{\ud\la'}{2\pi} h_n(\la') a_{nk} (\la'-\la)\nu_k\vartheta_k(\la)\,.
\ee
The sign $\pm$ in Eq. \eqref{eq:qdr} corresponds to particle and hole excitations.
Taking the derivative of the dressed charge with respect to the rapidity we find
\be
\pm {q^\text{dr}}'_m(\la) =  q'_m(\la) -  \sum_{k=0}^p \int\frac{\ud\la'}{2\pi} h_k(\la')a_{km}(\la'-\la)\,.
\ee
Using standard manipulations one can show that ${q^\text{dr}}'_m(\la_\pm)$ satisfies the integral equation
\be
\pm {q^\text{dr}_m}'(\la) = q'_m(\la) -  \sum_{k=0}^p \int\frac{\ud\la'}{2\pi} {q^\text{dr}_k}'(\la')\nu_k\vartheta_k(\la')a_{km}(\la'-\la)\,,
\ee
which can also be seen by comparing its iterative solution with that of Eq. \eqref{heq}. In particular, comparing with Eq. \eqref{rhoeq1} we see that for the dressed momentum
\be
{p_m^\text{dr}}'(\la) = -\nu_m2\pi \rhot_m(\la)\,.
\ee
The dressed [or effective] velocity of an elementary excitation is defined as
\be
v_m^\text{dr}(\la) = \frac{ {e_m^\text{dr}}'(\la)}{{p_m^\text{dr}}'(\la)}\,,
\ee
and it satisfies
\be
\label{veq}
-\nu_m\rhot_m(\la)v_m^\text{dr}(\la) = \frac{e'_m(\la)}{2\pi} + \sum_{k=0}^p \left(a_{mk}*(\rho_k v^\text{dr}_k)\right)(\la)\,.
\ee
We note that, using again Eq. \eqref{rhoeq1} on the left hand side of Eq. \eqref{veq}, we can write it as
\be
p'_m(\la)v_m^\text{dr}(\la) = p'_m(\la)v_m(\la) +  
\sum_{k=0}^p \int\frac{\ud\la'}{2\pi} a_{mk}(\la-\la')\rho_k(\la')\left[v^\text{dr}_k(\la')-v^\text{dr}_m(\la)\right],
\ee
which is the form written in Ref. \cite{CaDY16} for a single particle species.

\section{Technical details on the non-relativistic limit.}
\label{sec:appendix_double_scaling}

In this appendix we provide a derivation of the non-relativistic-limit formulae presented in Sec.~\ref{sec:sanity}.

\subsection{Non-relativistic limit of the Bethe--Takahashi equations}

Here we explicitly take the non-relativistic limit \eqref{eq:def_double_scaling} of the Bethe--Takahashi equations \eqref{eq:rhoeq}. Let us start by computing the limit of \eqref{eq:rhoeq} for $k=0$. Using the definition \eqref{eq:rhotk} we have 
\be
\tilde \rho^{(t)}_0(k)   = \frac{1}{2\pi}+\frac{1}{M c\cosh\theta(k) }\sum_{m=0}^p (a_{0m}*\rho_m)(\th(k))\,.
\label{eq:apprho0}
\ee
We now show that the second term on the r.h.s. is $O(1/c)$. This is achieved by showing that the every element of the sum is well defined in the non-relativistic limit. Let us consider the first 
\begin{align}
\limsc (a_{00}*\rho_0)(\th(k)) &= \limsc \int \frac{{\rm d} k'}{2\pi}a_{00}(\theta(k)-\theta(k'))\tilde\rho_0(k')\notag\\
&=\limsc\int \frac{{\rm d} k'}{2\pi} a_{00}\left(\frac{k-k'}{Mc}\right)\tilde\rho_0(k')\notag\\
&= a_{00}(0) \frac{n}{2\pi}\,.
\label{eq:applimconv}
\end{align}
Here we used that $\tilde\rho_0(k)$ has support on a finite region in the non-relativistic limit, so that \eqref{eq:fixedmomentum} applies. Instead, for $0<m\leq p$ we have 
\begin{align}
\limsc (a_{0m}*\rho_m)(\th(k)) &= \limsc \int \frac{{\rm d} \lambda}{2\pi}a_{0m}(\theta(k)-\lambda)\rho_m(\lambda)\notag\\
 &= \int \frac{{\rm d} \lambda}{2\pi}a_{0m}(\lambda)\rho_m(\lambda)\,.
 \label{eq:applimconv2}
\end{align}
Eqs. \eqref{eq:applimconv} and \eqref{eq:applimconv2} prove that the second term on the r.h.s. of \eqref{eq:apprho0} is indeed negligible in the non-relativistic limit and we find \eqref{eq:scrho}. Considering now \eqref{eq:rhoeq} for $1\leq k \leq p$, we have 
\be
\rho^t_k(\lambda) =\nu_k (a_{k}*\rho_0)(\lambda)-\nu_k\!\!\sum_{m=1}^p (a_{km}*\rho_m)(\lambda)\,,
\label{eq:apprhok}
\ee
Proceeding as in \eqref{eq:applimconv} we find 
\be
\limsc(a_{k}*\rho_0)(\lambda)= a_{k}(\lambda)  \frac{n}{2\pi}\,.
\ee
This shows that the non-relativistic limit of equation \eqref{eq:apprhok} is indeed \eqref{eq:scmag} and concludes the derivation.

\subsection{TBA in the non-relativistic limit}

In this subsection we write down and solve the thermodynamic Bethe Ansatz equations in the  non-relativistic limit~\eqref{eq:def_double_scaling}.  

\subsubsection{Non-relativistic limit of the thermodynamic Bethe Ansatz equations}

The first step is to take the non-relativistic limit of the thermodynamic Bethe Ansatz equations \eqref{eq:TBA}. This can be done by making two simple observations. First, we note that any convolution involving $\ln(1+ \eta_0^{-1})(\th)$ is negligible in the limit. This is because, by assumption,  $\tilde\rho_0(k)$ has support on a finite region in the non-relativistic limit, so that 
\be
\eta_0(\th(k)) =  \frac{\tilde\rho^h_0(k)}{\tilde \rho_0(k)} \equiv \tilde \eta_0(k)
\ee
is also supported on a finite region in the non-relativistic limit. Considering the convolution of $\ln(1+ \eta_0^{-1})(\th)$ with any smooth function $a(\th)$ we then find
\begin{align}
a*\ln(1+ \eta_0^{-1})(\th) &= \int \frac{{\rm d} k}{2\pi\sqrt{M^2 c^2+ k^2}}a(\th-\sinh^{-1}(k/(Mc)))\ln(1+\tilde \eta_0^{-1})(k)\notag\\
&=\frac{a(\th)}{ M c}\int \frac{{\rm d} k}{2 \pi} \ln(1+\tilde \eta_0^{-1})(k) + O\left(\frac{1}{c^2}\right)\,.
\end{align}
The second observation is that, since $a_{m0}(\th)$ are smooth functions of $\th$ and $k$ is fixed, any convolution ${a_{m0}*\ln(1+\eta_m^{-1})(\th(k))}$ becomes ${a_{m0}*\ln(1+\eta_m^{-1})(0)}$ in the non-relativistic limit. Using the two observations above and the definition 
\be
\limsc (\mu_1 - M c^2)=  \tilde \mu_1\,, 
\ee
from  \eqref{eq:TBA} we obtain 
\bes
\label{eq:applimTTBA}
\begin{align}
\ln\tilde\eta_0(k) &= \frac{ k^2/(2M)-\tilde \mu_1- \mu_2}{T} + \sum_{m=1}^p \nu_m a_{m0}*\ln(1+\eta_m^{-1})(0)\,,\label{eq:applimTTBA1}\\
\ln\eta_k &= \frac{2 \mu_2}{T}\ell_k +\sum_{m=1}^p \nu_m a_{mk}*\ln(1+\eta_m^{-1})\,.\label{eq:applimTTBA2}
\end{align}
\esu
Note that these equations could have been obtained by minimising the functional 
\be
\tilde g= \int\!\!{\rm d}k  \frac{k^2}{2M} \tilde \rho_0(k)-\tilde \mu_1 n - \mu_2 q - T s\,,
\label{eq:Gtilde}
\ee
and using the equations \eqref{eq:scBT}. The constant $\sum_{m=1}^p \nu_m a_{m0}*\ln(1+\eta_m^{-1})(0)$ on the r.h.s. of \eqref{eq:applimTTBA1} comes from the variation of $n$ in \eqref{eq:scmag}. 

Equations \eqref{eq:applimTTBA} can be further simplified by noting that, since the driving term of \eqref{eq:applimTTBA2} is constant, $\{\eta_k\}_{k=1}^p$ are independent of $\lambda$. Using 
\bes
\label{eq:appintas}
\begin{align}
\int\frac{{\rm d}\lambda}{2\pi}\,  a_{p0}(\lambda) =-\int\frac{{\rm d}\lambda}{2\pi}\, a_{1}^-(\lambda)&=\frac{1}{p}\,,\label{eq:appintam}\\
\int\frac{{\rm d}\lambda}{2\pi}\,  a_{m0}(\lambda)=- \int\frac{{\rm d}\lambda}{2\pi}\, a_{m}(\lambda)&=\frac{m-p}{p}\qquad \qquad\qquad m=1,\ldots,p-1, \label{eq:appinta}
\end{align}
\esu
we then find 
\bes
\label{eq:limTTBA}
\begin{align}
\ln\tilde\eta_{0}(k) &=\frac{ k^2/(2M)-\tilde \mu_1- \mu_2}{T} - \frac{1}{p} \ln(1+\eta_p^{-1}) -\!\!\sum_{m=1}^{p-1} \left[1-\frac m p\right]\ln(1+\eta_m^{-1})\,,\label{eq:limTTBA1}\\
 \ln\eta_{k} &= 2\frac{ \mu_2}{T}\ell_k + \sum_{m=1}^p \nu_m a_{mk}*\ln(1+\eta_m^{-1})\,.\label{eq:limTTBA2}
\end{align}
\esu
Here, for later convenience, we kept the convolution in the second equation.

\subsubsection{Solution of Eqs. \eqref{eq:limTTBA}}
\label{app:solNRTBA}

Equations \eqref{eq:limTTBA} can be solved exactly. The explicit solution can be found as follows. First we note that \eqref{eq:limTTBA2} are the TBA equations of a gapless XXZ spin-1/2 chain at infinite temperature and finite magnetisation, their solution reads as~\cite{Taka} 
\bes
\begin{align}
&\eta_{k}(h) = \!\!\left[\frac{\sinh[(k+1)h ]}{\sinh(h)}\right]^2\!\!\!-1\,,\,\,\quad1\leq k<p-1\,,\\
&\eta_{p-1}(h)= e^{ph}\frac{\sinh[(p-1)h]}{\sinh(h)}\,,\\
&\eta_{p}(h) = e^{p h}\frac{\sinh(h)}{\sinh[(p-1)h]}\,,
\end{align}
\label{eq:solution_etas}
\esu
where $h=\mu_2/T$. Using the identity 
\begin{align}
\sum_{m=1}^{p-1}\left[1-\frac m p\right]\ln(1+\eta_m^{-1}(h))+ \frac 1 p\ln(1+\eta_p^{-1}(h))=\log(1+e^{-2h})
\label{eq:appsumetas}
\end{align}
we finally obtain 
\be
\tilde \eta_{0}(k)= \frac{e^{\frac{ k^2/(2M)-\tilde \mu_1}{T}}}{2\cosh({\mu_2}/{T})}\,.
\label{eq:solution_eta_0}
\ee

\subsubsection{Solution of Eqs. \eqref{eq:scBT}}

The analytic expressions \eqref{eq:solution_etas} and \eqref{eq:solution_eta_0} can now be plugged into the linear equations \eqref{eq:scBT}, obtaining a closed system for the functions $\tilde{\rho}_0(k)$ and $\rho_n(k)$. Remarkably also this system can be solved exactly as we now show. First, plugging \eqref{eq:solution_eta_0} in \eqref{eq:limTTBA1} we find 
\be
\tilde\rho_0(k) =\frac{1}{2\pi}\left[1+\frac{\exp[{\frac{k^2}{2MT}-\frac{\tilde \mu_{1}}{T}}]}{2\cosh(\frac{\mu_{2}}{T})}\right]^{-1}\,.
\label{eq:thermalrho0sc}
\ee
Next, we consider the decoupled form of \eqref{eq:scmag}  
\bes
\label{eq:decoupledlowT}
\begin{align}
& \rhot_{1}(\th)=\frac{1}{2\pi} r(\th) n+r\ast\rhoh_{2}(\th)\,,\label{eq:decoupledlowT1}\\
& \rhot_k(\th) =r\ast\left(\rhoh_{k-1}+\rhoh_{k+1}\right)(\th)\,, & 2 \leq k \leq p-3\,,\label{eq:decoupledlowT2}\\
& \rhot_{p-2}(\th) =r\ast\left(\rhoh_{p-3}+\rhoh_{p-1}+\rho_{p}\right)(\th)\,, \label{eq:decoupledlowT3}\\
&\rhot_{p}(\th) =\rhot_{p-1}(\th)= r\ast\rho^{(\rm h)}_{p-2}(\theta)\,,
\label{eq:decoupledlowT4}
\end{align}
\esu
where $n$ is the particle density in the thermal state and we introduced 
\be
r(x)\equiv\int_{-\infty}^{\infty} {\rm d}y\,\frac{e^{-i x y}}{2\cosh\left(\frac{\pi y}{2}\right)}\,.
\label{eq:rfunction}
\ee
We then define 
\bes
\begin{align}
&\mathcal F[\rhoh_k](x)=f_k(x)\,,& &k =0,\ldots,p-2\,,\\
&\mathcal F[\rhoh_{p-1}+\rho_p](x)=f_{p-1}(x)
\end{align}
\esu
such that $f_k(x)$ fulfil the recurrence equation 
\be
(e^{\frac{\pi}{2}x}+e^{-\frac{\pi}{2}x})\alpha_k f_k(x) =f_{k-1}(x)+f_{k+1}(x)\,,
\label{eq:genericsystem}
\ee
where 
\be
\alpha_k=\frac{\sinh^2[(k+1)h]}{\sinh[(k+2)h]\sinh[kh]}\,,
\ee
and, to lighten notation, we set $h=\mu_2/T$. The boundary conditions on $f_k(x)$ are given by 
\bes
\begin{align}
f_0(x)&=n\,, \\
2\cosh\left(\frac{\pi x}{2}\right) \sinh(h p)  f_{p-1}(x)&= 2\cosh(h)\sinh(h (p-1)) f_{p-2}(x)\,.
\end{align}
\esu
The system \eqref{eq:genericsystem} coincides with Eq. 8.62 in Chapter 8.4.2 of \cite{Taka} and has the following general solution
\begin{align}
f_{k}(x)=& A(x) \left[\frac{\sinh[(k+2)h]}{\sinh[(k+1)h]} e^{-\frac{\pi}{2}k|x|}-\frac{\sinh[kh]}{\sinh[(k+1)h]} e^{-\frac{\pi}{2}(k+2)|x|}\right]\notag\\
&+B(x) \left[\frac{\sinh[(k+2)h]}{\sinh[(k+1)h]} e^{\frac{\pi}{2}k|x|}-\frac{\sinh[kh]}{\sinh[(k+1)h]} e^{\frac{\pi}{2}(k+2)|x|}\right]\,.
\label{eq:generalsolution}
\end{align}
Imposing the boundary conditions we find 
\be
f_{k}(x)=\frac{n}{2 \cosh(h)} \left(\frac{\sinh[(k + 2) h]}{\sinh[(k + 1) h]} \frac{\sinh[(p - k) \frac{\pi x}{2}]}{\sinh[p \frac{\pi x}{2}]} -\frac{\sinh[k h]}{\sinh[(k + 1) h]} \frac{\sinh[(p-k-2) \frac{\pi x}{2}]}{\sinh[p \frac{\pi x}{2}]}\right),
\ee
where $k=0,\ldots,p-1$. Taking the inverse Fourier transform and using \eqref{eq:decoupledlowT4} we finally obtain 
\bes
\label{eq:apprhos}
\begin{align}
{\rho}_{k}(\lambda,h,n)&=\frac{n \tanh(h)\sinh(h)}{4\pi\sinh[(k+1) h]} \left[\frac{ a_k(\lambda)}{\sinh[kh]} -\frac{ a_{k+2}(\lambda)}{\sinh[(k+2) h]} \right]\,, \qquad\qquad\qquad 1 \leq k \leq p-2\,,\label{eq:apprhos1}\\
{\rho}_{p-1}(\lambda,h,n)&=\frac{n \tanh[h]\left(\sinh[h]+e^{-p h}\sinh[(p-1)h]\right)}{4\pi\sinh[(p-1) h]\sinh[p h]}{a_{p-1}(\lambda)}\,,\label{eq:apprhos2}\\
{\rho}_{p}(\lambda,h,n)&=\frac{n \left(e^{-p h}\sinh[h]+\sinh[(p-1)h]\right)}{4\pi\cosh[h]\sinh[p h]}{a_{p-1}(\lambda)}\,. \label{eq:apprhos3}
\end{align}
\esu

\subsubsection{Expectation value of topological charge}

Given the analytic knowledge of $\tilde{\rho}_0(k)$ and $\rho_n(\lambda)$ we can determine all the charge densities, which are then simple integrals. In the case of the topological charge, however, these integrals can be further simplified. To proceed we consider the functional $\tilde g$ in Eq.~\eqref{eq:Gtilde}. By means of the identity Eq.~\eqref{eq:identity} we readily obtain
\be
\tilde g =- T\int\frac{\ud k}{2\pi} \ln\left(1+2 \cosh(\mu_2  \beta) e^{- \beta (k^2/(2M)-\tilde\mu_1)}\right)\,.
\ee
Note that 
\be
\ud\tilde g=-s\,\ud T-n \,\ud\mu_1-q \ud\mu_2\,,
\ee
so we have 
\bes
\begin{align}
&s=\frac{1}{ T^2 }\frac{\partial{\tilde g}}{\partial  \beta}\Big |_{\tilde \mu_1,\mu_2}\,,\label{eq:thermaS}\\
&n=-\frac{\partial{\tilde g}}{\partial \tilde \mu_1}\Big |_{\beta,\mu_2}=\int\frac{\ud k}{2\pi} \frac{2 \cosh(\mu_2  \beta)}{2 \cosh(\mu_2  \beta)+e^{ \beta (k^2/(2M)-\tilde\mu_1)}} = \int {\ud k} \tilde \rho_0(k)\,,\label{eq:thermaln}\\
&q=-\frac{\partial{\tilde g}}{\partial  \mu_2}\Big |_{\beta,\tilde\mu_1}=\int\frac{\ud k}{2\pi} \frac{2 \sinh(\mu_2  \beta)}{2 \cosh(\mu_2  \beta)+e^{ \beta (k^2/(2M)-\tilde\mu_1)}}=\tanh(\mu_{2}  \beta) n\,, \label{eq:derivation_thermaltopcharge}
\end{align}
\esu
Note that the last equation implies the non-trivial identity
\be
\frac{1}{n}\sum_{m=1}^p  \ell_m\int\!\ud\th\,  \rho_m(\th,h,n)=\frac{e^{-h}}{2 \cosh(h)}\,,
\label{eq:nontrivialidentity}
\ee
where $\{\rho_m(\th,h)\}$ are those in Eqs. \eqref{eq:apprhos}. We could not prove this identity analytically but we verified it numerically to arbitrary precision.

\section{Technical details on  the small-temperature limit.}
\label{sec:appendix_small_T_limit}

In this appendix we provide a derivation of the formulae presented in Sec.~\ref{sec:lowT}. The starting point is to take the small temperature limit of \eqref{eq:TBA}. Assuming 
\be
\eta_{0}=O(e^{A\beta}),\qquad A>0,\qquad\qquad \eta_{n}=O(\beta^0),\qquad 1<n<p\,,
\label{eq:appass}
\ee
we can neglect all terms containing $\ln(1+\eta_0^{-1})(\th)$ and the equations completely decouple, yielding
\bes
\begin{align}
\ln\eta_{0}(\theta) &= \beta(M  \cosh\th-\mu_1)-h - \frac{1}{p}\sum_{m=1}^{p-1} (p-m) \ln(1+\eta_m^{-1})- \frac{1}{p} \ln(1+\eta_p^{-1})+O(e^{-A\beta})\,,\label{eq:etanlowt1}\\
 \ln\eta_{n} &= \frac{2h}{\ell_n} + \sum_{m=1}^p \nu_m a_{mn}*\ln(1+\eta_m^{-1})+O(e^{-A\beta}), \qquad\qquad\qquad\qquad\qquad n=1,\dots,p\,,\label{eq:etanlowt2}
\end{align}
\esu
where we used the identities \eqref{eq:appintas}. Equations \eqref{eq:etanlowt2} are solved in Appendix~\ref{app:solNRTBA}. Using the explicit solution \eqref{eq:solution_etas} and the identity \eqref{eq:appsumetas} we then find 
\be
\eta_{0}(\th)=\frac{e^{\beta (Mc^2\cosh\th-\mu_1)}}{2\cosh h}\,.
\label{eq:eta0smallt}
\ee
Note that the solution is consistent with the assumption \eqref{eq:appass}. Plugging these expressions in \eqref{eq:continuity}--\eqref{eq:vrhoray} and neglecting subleading corrections we have    
\bes
\label{eq:appetalowT}
\begin{align}
\eta_{0,\zeta}(\lambda) &= \eta_{0, \rm L}(\lambda)\Theta[v_{0,\zeta}(\lambda)-\zeta] + \eta_{0, \rm R}(\lambda) \Theta[\zeta-v_{0,\zeta}(\lambda)]\,,\label{eq:appetalowT1}\\
\eta_{n,\zeta}(\lambda) &= \eta_{n}(h_{\rm L})\Theta[v_{n,\zeta}(\lambda)-\zeta] + \eta_{n}(h_{\rm R})\Theta[\zeta-v_{n,\zeta}(\lambda)]\,,\label{eq:appetalowT2}
\end{align}
\esu
\bes
\label{eq:rholt}
\begin{align}
\rhot_{0,\zeta}(\lambda)  &= \frac{1}{2\pi} {Mc\cosh\lambda}+\sum_{m=0}^p (a_{0m}*\frac{1}{1+\eta_{m,\zeta}}\rhot_{m,\zeta})(\lambda)\,,\label{eq:rho0lt}\\
\rhot_{k,\zeta}(\lambda)  &= -\nu_k \sum_{m=0}^p (a_{km}*\frac{1}{1+\eta_{m,\zeta}}\rhot_{m,\zeta})(\lambda)\,,  &1\leq k \leq p\,, \label{eq:rhonlt}
\end{align}
\esu
\bes
\label{eq:vlt}
\begin{align}
\rhot_{0,\zeta}(\lambda)  v^{\rm dr}_{0,\zeta}(\lambda)  &= \frac{1}{2\pi}{Mc^2\sinh\lambda}+\sum_{m=0}^p (a_{0m}*\frac{1}{1+\eta_{m,\zeta}}\rhot_{m,\zeta}v^{\rm dr}_{m,\zeta})(\lambda)\,,\label{eq:v0lt}\\
\rhot_{k,\zeta}(\lambda)  v^{\rm dr}_{k,\zeta}(\lambda)  &=  - \nu_k \sum_{m=0}^p (a_{k m}*\frac{1}{1+\eta_{m,\zeta}}\rhot_{m,\zeta} v^{\rm dr}_{m,\zeta})(\lambda)\,,  &1\leq k \leq p\,. \label{eq:vnlt}
\end{align}
\esu

Here $\eta_{0, \rm L/\rm R}(\lambda)$ are of the form \eqref{eq:eta0smallt} where the parameters have subscripts ${\rm L}$ and ${\rm R}$. To simplify \eqref{eq:rholt}--\eqref{eq:vlt} we make the following assumptions
\bes
\begin{align}
&\rhot_{0,\zeta}(\lambda)=O(1)\,,\label{eq:ass1}\\
&\rhot_{k,\zeta}(\lambda)=O\left(\beta^{-1/2}e^{-M c^2 \beta}\right)\,, \qquad 1\leq k \leq p\,.
\label{eq:ass2}
\end{align}
\esu
At the end of the calculation we will verify the consistency of these assumptions. Using these assumptions we have
\begin{align}
\rhot_{0,\zeta}(\lambda)&=\frac{1}{2\pi}{Mc\cosh\lambda}\,,\label{eq:lowTrhot}\\ 
v^{\rm dr}_{0,\zeta}(\lambda)\rhot_{0,\zeta}(\lambda)&=\frac{1}{2\pi}{Mc^2\sinh\lambda}\,,
\end{align}
so that we can write
\be
v^{\rm dr}_{0,\zeta}(\lambda) = \frac{\rhot_{0,\zeta}(\lambda)  v_{0,\zeta}(\lambda)}{\rhot_{0,\zeta}(\lambda)}= c\,{\tanh\lambda}\,,
\ee
and thus
\begin{align}
\eta_{0,\zeta}(\lambda) =  \frac{e^{ \beta_{\rm L} (Mc^2\cosh\th-\mu_{1,\rm L})}}{2\cosh(h_{\rm L})}\Theta[{\tanh\lambda}-\zeta/c] +\frac{e^{\beta_{\rm R}(Mc^2 \cosh\th-\mu_{1,\rm R})}}{2\cosh(h_{\rm R})}\Theta[\zeta/c-{\tanh\lambda}]\,,
\end{align}
and 
\begin{align}
\rho_{0,\zeta}(\lambda)=&\frac{Mc\cosh\lambda\cosh(h_{\rm L})}{\pi\left[2\cosh(h_{\rm L})+e^{\beta(Mc^2\cosh\theta-\mu_{1,\rm L})}\right]}\Theta[{\tanh\lambda}-\zeta/c]\notag\\
&+\frac{Mc\cosh\lambda\cosh(h_{\rm R})}{\pi\left[2\cosh(h_{\rm R})+e^{\beta(Mc^2\cosh\theta-\mu_{1,\rm R})}\right]}\Theta[\zeta/c-{\tanh\lambda}].
\end{align}
Plugging back into \eqref{eq:rhonlt} and \eqref{eq:vnlt} we have 
\begin{align}
\rhot_{k,\zeta}(\lambda)  &= \nu_k (a_{k}*\frac{1}{1+\eta_{0,\zeta}}\rhot_{0,\zeta})(\lambda)-\nu_k\sum_{m=1}^p (a_{km}*\frac{1}{1+\eta_{m,\zeta}}\rhot_{m,\zeta})(\lambda)\,,  & 1\leq k \leq p\,,\label{eq:rhonltstep1}\\
\rhot_{k,\zeta}(\lambda)  v^{\rm dr}_{k,\zeta}(\lambda)  &=  \nu_k (a_{k}*\frac{1}{1+\eta_{0,\zeta}}\rhot_{0,\zeta} v^{\rm dr}_{0,\zeta})(\lambda) - \nu_k \sum_{m=1}^p (a_{k m}*\frac{1}{1+\eta_{m,\zeta}}\rhot_{m,\zeta} v^{\rm dr}_{m,\zeta})(\lambda)\,,  & 1\leq k \leq p\,.
\end{align}
Let us consider the driving term of \eqref{eq:rhonltstep1}. Writing it explicitly we have 
\begin{align}
&\nu_k (a_{k}*\frac{1}{1+\eta_{0,\zeta}}\rhot_{0,\zeta})(\lambda)=\int\frac{{\rm d}\mu}{2\pi}  \nu_k a_{k}(\lambda-\mu) \rho_{0,\zeta}(\mu)= \frac{1}{2\pi}\nu_k a_{k}(\lambda) n(\zeta)+O(\beta^{-1})\,,
\end{align}
where we introduced 
\begin{align}
n(\zeta)&=\int{{\rm d}\mu}\,\rho_{0,\zeta}(\mu)\notag\\
&=\int\frac{{{\rm d}\mu}}{2\pi} \left(\frac{2\cosh h_{\rm L}M c \cosh\mu}{2\cosh h_{\rm L}+e^{\beta_{\rm L}(Mc^2\cosh\theta-\mu_{1,\rm L})}}\Theta[{\tanh\mu}-\zeta/c] +\frac{2\cosh h_{\rm R} M c \cosh\mu}{2\cosh h_{\rm R}+e^{\beta_{\rm R}(Mc^2\cosh\theta-\mu_{1,\rm R})}}\Theta[\zeta/c-{\tanh\mu}]\right)\notag\\
&=\sqrt{\frac{M}{2\pi \beta}}\left( e^{-\Delta_{\rm L}\beta}\cosh h_{\rm L} {\rm erfc}\left(\sqrt{\frac{M \beta_{\rm L}}{2}}\zeta\right)+  \cosh h_{\rm R} e^{-\Delta_{\rm R}\beta} e^{-M\Delta\beta} {\rm erfc}\left(-\sqrt{\frac{M \beta_{\rm R}}{2}}\zeta\right)\right)(1+O(\beta^{-1/2}))
\label{eq:lowTdensity}
\end{align}
with  $\Delta_{\rm L/R}=Mc^2-\mu_{1,\rm L/R}$. Analogously, we have 
\begin{align}
\nu_k (a_{k}*\frac{1}{1+\eta_{0,\zeta}}v^{\rm dr}_{0,\zeta}\rhot_{0,\zeta})(\lambda)=\int\frac{{\rm d}\mu}{2\pi}  \nu_k a_{k}(\lambda-\mu) v^{\rm dr}_{0,\zeta}(\mu)\rho_{0,\zeta}(\mu)= \frac{1}{2\pi}\nu_k a_{k}(\lambda) J_n(\zeta)+O(\beta^{-1})\,,
\end{align}
where 
\begin{align}
J_n(\zeta)&=\int{{\rm d}\mu}\,v^{\rm dr}_{0,\zeta}(\mu)\rho_{0,\zeta}(\mu)\notag\\
&=\int{{\rm d}\mu} \left(\frac{2\cosh h_{\rm L}M c^2 \sinh\mu}{2\cosh h_{\rm L}+e^{\beta_{\rm L}(Mc^2\cosh\theta-\mu_{1,\rm L})}}\Theta[{\tanh\mu}-\zeta/c] +\frac{2\cosh h_{\rm R} M c^2 \sinh\mu}{2\cosh h_{\rm R}+e^{\beta_{\rm R}(Mc^2\cosh\theta-\mu_{1,\rm R})}}\Theta[\zeta/c-{\tanh\mu}]\right)\notag\\
&=\frac{c}{\pi \beta}\left( e^{-\Delta_{\rm L}\beta} \cosh h_{\rm L} e^{-\frac{M \beta_{\rm L}}{2}\zeta^2}-  \cosh h_{\rm R}e^{-\Delta_{\rm R}\beta} e^{-M \Delta\beta} e^{-\frac{M \beta_{\rm R}}{2}\zeta^2}\right) (1+O(\beta^{-1/2})).
\end{align}
Putting all together, at the leading order, we find 
\begin{align}
\rhot_{k,\zeta}(\lambda)  &= \frac{1}{2\pi}\nu_k a_{k}(\lambda) n(\zeta)-\nu_k \sum_{m=1}^p (a_{km}*\frac{1}{1+\eta_{m,\zeta}}\rhot_{m,\zeta})(\lambda)\,,  & 1\leq k \leq p\,,\\
\rhot_{k,\zeta}(\lambda)  v^{\rm dr}_{k,\zeta}(\lambda)  &= \frac{1}{2\pi}\nu_k a_{k}(\lambda) J_n(\zeta) - \nu_k \sum_{m=1}^p (a_{k m}*\frac{1}{1+\eta_{m,\zeta}}\rhot_{m,\zeta} v^{\rm dr}_{m,\zeta})(\lambda)\,,  &  1\leq k \leq p\,.
\end{align}
Since the driving terms are proportional we immediately find 
\be
v^{\rm dr}_{n,\zeta}(\lambda)=\frac{\rhot_{n,\zeta}(\lambda)  v^{\rm dr}_{n,\zeta}(\lambda)}{\rhot_{n,\zeta}(\lambda)}=\frac{J_n(\zeta)}{n(\zeta)}\equiv v(\zeta)\,.
\ee
Plugging it back in Eqs. \eqref{eq:appetalowT} and following the steps outlined in Sec.~\ref{sec:sanity} we then find
\be
\rho_{k,\zeta}(\lambda) = \frac{n(\zeta)}{n_{\rm L}}\rho_{k,\rm L}(\lambda)\Theta[v({\zeta})-\zeta] +\frac{n(\zeta)}{n_{\rm R}}\rho_{k,\rm R}(\lambda)\Theta[\zeta-v({\zeta})]\,,
\ee
where the root densities $\{\rho_{m, {\rm L/R}}(\th)\}$ solve \eqref{eq:decoupledlowT}. We can then use the solutions \eqref{eq:apprhos} (respectively setting $h=h_{\rm L/R}$). 

Finally, we note that the assumption \eqref{eq:ass1} is verified, as demonstrated by \eqref{eq:lowTrhot}. Moreover, as reported in \eqref{eq:apprhos},  $\{\rho_{k,\zeta}(\lambda)\}_{k=1}^p$ are proportional to the density of particles. Using that $n=O(\beta^{-1/2}e^{-M c^2 \beta})$ [cf. \eqref{eq:lowTdensity}] we have that the assumption \eqref{eq:ass2} is also verified. This proves the self-consistency of our solution.

\twocolumngrid

\end{document}